\documentclass[aip,pof,reprint,onecolumn]{revtex4-1}

\usepackage{amsmath,bm,amssymb,psfrag,color,float}
\usepackage{graphicx}

\newcommand{\C}{\mathcal{C}}
\newcommand{\tC}{\mathcal{C}^{\mathrm{P}}}
\newcommand{\tT}{\mathcal{T}^{\mathrm{P}}}
\newcommand{\p}{\mathrm{P}}
\newcommand{\Wi}{\mathrm{Wi}}
\newcommand{\tr}{\operatorname{tr}}

\begin{document}

\title{On the Peterlin approximation for turbulent flows
of polymer solutions}
\author{Dario Vincenzi}
\affiliation{Laboratoire Jean Alexandre Dieudonn\'e, Universit\'e Nice Sophia Antipolis, CNRS, UMR 7351, 06100 Nice, France}
\author{Prasad Perlekar}
\affiliation{TIFR Center for Interdisciplinary Sciences, Tata Institute of
Fundamental Research, Narsingi, Hyderabad 500075, India}
\author{Luca Biferale}
\affiliation{Department of Physics and INFN, University of Rome 
``Tor Vergata'', Via della Ricerca Scientifica 1, 00133 Roma, Italy}
\author{Federico Toschi}
\affiliation{Department of Physics, and Department of Mathematics and Computer Science, Eindhoven University of Technology, 5600 MB Eindhoven, The Netherlands\\
IAC, CNR, Via dei Taurini 19, I-00185 Roma, Italy}
\date{\today}

\begin{abstract}
We study the impact of the Peterlin approximation on the
statistics of the end-to-end separation of polymers in a turbulent
flow.
The FENE and FENE-P models are numerically integrated along a
large number of Lagrangian trajectories resulting from a direct
numerical simulation of three-dimensional homogeneous isotropic
turbulence.  Although the FENE-P model yields results in qualitative
agreement with those of the FENE model, quantitative differences
emerge.  The steady-state probability of large extensions is
overestimated by the FENE-P model.  The alignment of polymers with the
eigenvectors of the rate-of-strain tensor and with the direction of
vorticity is weaker when the Peterlin approximation is used.  At large
Weissenberg numbers, both the correlation times of the extension and
of the orientation of polymers are underestimated by the FENE-P model.
\end{abstract}


\maketitle

\section{Introduction}

The addition of elastic polymers to a Newtonian solvent
introduces a history dependence in the response of the fluid to 
a deformation and hence modifies the rheological properties of the
solvent.~\cite{BHAC77}
In turbulent flows, the non-Newtonian nature of polymer solutions
manifests itself through a considerable reduction of the turbulent
drag compared to that of the solvent alone.~\cite{BLP08,WM08,PPR09,B10,G14}
What renders this phenomenon
even more remarkable is that an appreciable drag reduction can already be
observed at very small polymer concentrations
(of the order of a few parts per million).
Turbulent drag reduction was 
discovered by Toms~\cite{T49} more than sixty years ago and
is nowadays routinely used to reduce energy losses in crude-oil
pipelines.~\cite{GB95}
A full understanding of turbulent drag reduction nonetheless remains 
a difficult challenge, because in the turbulent flow of a polymer solution 
the extensional dynamics of a large number of 
polymers is coupled
with strongly nonlinear transfers of kinetic energy.

The study of turbulent flows of polymer solutions
is essentially based on two approaches: the molecular approach and the
continuum one.
In the molecular (or Brownian Dynamics) 
approach, a polymer is modeled as a sequence
of~$N$ beads connected by elastic springs.  
The deformation of the bead-spring chain is then followed 
along the trajectory of its center of mass.
In homogeneous and isotropic turbulence,
Watanabe and Gotoh~\cite{WG10} have shown that~$N=2$ beads are in fact sufficient
to describe the stationary statistics of both the extension and the orientation
of polymers, i.e., the deformation of a polymeric chain is dominated
by its slowest oscillation mode. An analogous conclusion has been reached by Terrapon
et al.~\cite{TDMS03} in a study of polymer dynamics in a turbulent channel flow. 
The model consisting of only~$N=2$ beads is known as the finitely 
extensible nonlinear elastic (FENE) dumbbell model.~\cite{BHAC77}
The molecular approach is suitable for
studying the deformation of passively transported
polymers \cite{MKSH93,IDKCS02,ZA03,TDMS03,
TDMSL04,GSK04,CMV05,CPV06,DS06,JC07,WG10,MV11,BMPB12}. Two-way coupling molecular
simulations are a recent achievement~\cite{PS07,WG13,WG14} and
have not yet been employed in practical applications owing to their computational cost. 
In the molecular approach, indeed,
the feedback of polymers on the velocity field is
given by the forces exerted on the fluid
by a huge number of polymeric chains.~\cite{PS07,WG13,WG14}

For the above reason, in practical applications,
numerical simulations of turbulent flows of polymer solutions generally
use the continuum approach. The conformation of polymers
is then described by means of a space- and time-dependent 
tensor field, which represents
the average inertia tensor of polymers at a given time and position in the 
fluid. Such tensor is termed the
polymer conformation tensor.
An evolution equation for the conformation tensor may in principle be 
derived from the FENE dumbbell model. Such equation, however, 
involves the average over thermal fluctuations
of a nonlinear function of the polymer 
end-to-end vector; a closure approximation is therefore required.
Peterlin~\cite{P66} proposed
a mean-field closure according to which the average of the elastic
force over thermal fluctuations is replaced by the value of
the force at the mean-squared polymer extension.
The resulting model was subsequently dubbed the
FENE-P model.\cite{BDJ80}
Within the FENE-P model,
the back-reaction of polymers on the flow is described by
a stress tensor field, which depends on the polymer conformation tensor.
This continuum model is thus suitable for simulating
turbulent flows
of polymer solutions; it indeed amounts to simultaneously
solving the evolution equation for the polymer conformation tensor and the Navier--Stokes equation
with an additional elastic-stress term.
The FENE-P model is widely employed
in numerical simulations of turbulent drag reduction and
has been successfully applied to channel flows, 
\cite{SBH97,DCP02,PBNHVH03,DWTSML04} 
shear flows, \cite{SRLWG04,RVCB10} and
two- and three-dimensional
homogeneous and isotropic turbulence.\cite{DCBP05,PMP06,PMP10,GPP12}
Nevertheless, although it qualitatively reproduces the main features of
turbulent drag reduction, the FENE-P model generally does not yield
results in quantitative agreement with experimental data
(e.g. Ref.~\onlinecite{DWTSML04}).
It is therefore essential to assess the validity of the 
assumptions on which the model is based.

For laminar flows, the Peterlin approximation has been examined in
detail (see Refs.~\onlinecite{H097,K97} and references therein). In particular,
the FENE-P model is
a good approximation of the FENE model in 
steady flows, whereas appreciable differences appear in time-dependent flows.
This observation suggests that in turbulent flows 
important differences  between the two models should be expected. Several studies have subsequently
investigated the validity of the Peterlin approximation in turbulent flows by comparing one-way coupling simulations of the FENE
and FENE-P models.\cite{IDKCS02, ZA03, TDMS03, GSK04,JC07}
These studies have clearly shown potential differences among FENE and FENE-P models together with high sensitivity on the
statistical ensemble and dependency on the degree of homogeneity of the underlying velocity fluctuations.
We undertake a systematic analysis of the Peterlin approximation 
in three-dimensional homogeneous and isotropic turbulence by means of one-way coupling Lagrangian
simulations of the FENE and FENE-P models.
The size of our statistical ensemble
($128\times 10^3$ fluid trajectories and~$2\times 10^3$
realizations of thermal noise per trajectory) allows us to fully characterize
the statistics of polymer extension and orientation.
When the flow is turbulent
two independent effects are at the origin of the discrepancies between the FENE and
FENE-P models: one is directly related to the closure
approximation for the elastic force, while the other is of a statistical
nature and is a consequence of deriving the statistics of polymer 
deformation from that of the conformation tensor. By isolating these two
effects, we compare the steady-state statistics and the temporal correlation of the
extension and of the orientation of polymers in the FENE and in the FENE-P model.

The rest of the paper is organized as follows.  In
Sect.~\ref{sec:models} we briefly review the FENE and FENE-P models.
The Lagrangian simulations are described in
Sect.~\ref{sec:Lagrangian}. The results of the simulations are
presented in Sect.~\ref{sec:results}. Finally, conclusions are drawn
in Sect.~\ref{sec:conclusions}.

\section{FENE and FENE-P models}
\label{sec:models}
In the FENE model, a polymer is described as two beads connected by
an elastic spring, i.e., as an elastic dumbbell.~\cite{BHAC77}
If the fluid is at rest, the polymer is in a coiled configuration 
because of entropic forces
and its equilibrium extension is determined by the intensity
of thermal fluctuations.
If the polymer is introduced in a moving fluid and the velocity field changes
over the size of the polymer, then the polymer can stretch and deform.
The dynamics of the polymer thus results from the interplay between the
stretching action of the velocity gradient and the elastic force, which
tends to take the polymer back to its equilibrium configuration.

The maximum extension of the dumbbell is assumed to be smaller than the
Kolmogorov scale, so that the velocity field changes
linearly in space at the scale of the dumbbell. The drag force on the beads
is given by the Stokes law. Moreover, inertial effects and 
hydrodynamical interactions between the beads are disregarded.
Polymer--polymer hydrodynamical interactions are also disregarded
under the assumption that the polymer concentration is very low.
Thus, the separation vector
between the beads, $\bm R(t)$, satisfies the following
stochastic ordinary differential equation (the FENE equation):~\cite{BHAC77,O96}
\begin{equation}
\label{eq:dumbbell}
\dfrac{d\bm{R}}{dt}=\sigma(t)\bm{R}-\dfrac{f\big(R^2\big)}{2\tau_p}\bm{R}
+\sqrt{\dfrac{R_0^2}{\tau_p}}\,\boldsymbol{\xi}(t),
\end{equation}
where $R=\vert\bm R\vert$, $\sigma_{ij}(t)=\partial_j u_i(t)$ is the velocity gradient at the
position of the center of mass of the dumbbell, 
$\bm\xi(t)$ is three-dimensional white noise,
$R_0$ is the polymer root-mean-square equilibrium extension,
and $\tau_p$ is the polymer relaxation time ($\tau_p$ is the time scale that describes the
exponential relaxation of $\langle R^2(t)\rangle$ to its equilibrium value
in the absence of flow). 
The three terms on the right-hand-side of Eq.~\eqref{eq:dumbbell} represent
the stretching by the velocity gradient, the restoring elastic force, and thermal noise, respectively.
The function~$f$ determines the elastic force and, in the FENE model, 
it has the following form:
\begin{equation}
\label{eq:elastic-force}
f(\zeta)=\dfrac{1}{1-\zeta/L^2},
\end{equation}
where $L$ is the maximum extension of the polymer.
The elastic force diverges as~$R$ approaches~$L$;
hence extensions greater than~$L$ are forbidden.
Note that~$\bm R(t)$ is a random vector and that, when~$\bm u$ is turbulent,
two independent sources of randomness 
influence its evolution: thermal noise and
the velocity gradient itself. 

The polymer conformation tensor is defined as 
$\mathcal{C}_{ij}\equiv\langle R_i R_j\rangle_\xi$,
where~$\langle\cdot\rangle_\xi$ denotes an average over thermal
fluctuations.
To derive the evolution equation for~$\mathcal{C}$, we apply the It\^o formula
to $R_iR_j$ and use  Eq.~\eqref{eq:dumbbell}: 
\begin{equation}
\label{eq:multiplication}
\dfrac{d}{dt}(R_iR_j)=\dfrac{dR_i}{dt} R_j+R_i\dfrac{dR_j}{dt}
+\dfrac{R_0^2}{\tau_p}\delta_{ij}=
\sigma_{ik}(t)R_kR_j+\sigma_{jk}(t)R_kR_i-\frac{1}{\tau_p}f(R^2)R_iR_j+
\dfrac{R_0^2}{\tau_p}\delta_{ij}+
\sqrt{\dfrac{R_0^2}{\tau_p}}\left[R_j\xi_i(t)+R_i\xi_j(t)\right].
\end{equation}
Whereas there is no It\^o--Stratonovich ambiguity for Eq.~\eqref{eq:dumbbell},
Eq.~\eqref{eq:multiplication} should be understood in the It\^o sense.
We now average Eq.~\eqref{eq:multiplication} with respect to the realizations
of~$\bm \xi(t)$  and make use of the following property of the
It\^o integral: $\langle R_i\xi_j(t)\rangle_\xi=0$.
We thus obtain:
\begin{equation}
\label{eq:non-closed}
\dfrac{d}{dt}\langle R_iR_j\rangle_\xi=
\sigma_{ik}(t)\langle R_kR_j\rangle_\xi+\sigma_{jk}(t)\langle R_kR_i\rangle_\xi
-\dfrac{1}{\tau_p}\left[\langle f(R^2)R_iR_j\rangle_\xi-R_0^2\delta_{ij}\right].
\end{equation}
Equation~\eqref{eq:non-closed} is not closed 
with respect to $\mathcal{C}$ because of the term:
\begin{equation}
\mathcal{A}_{ij}\equiv\big\langle f\big(R^2\big)R_iR_j\big\rangle_\xi.
\end{equation} 
To obtain a closed equation, 
Peterlin~\cite{P66} proposed the following approximation:
\begin{equation}
\label{eq:peterlin-closure}
\mathcal{A}_{ij}
\approx f\big(\big\langle R^2\big\rangle_\xi\big)
\big\langle R_iR_j\big\rangle_\xi=f(\operatorname{tr}\mathcal{C})
\mathcal{C}_{ij}.
\end{equation}
The resulting evolution equation for the polymer conformation tensor 
(the FENE-P equation) is:
\begin{equation}
\label{eq:FENE-P}
\dfrac{d\tC}{dt}=
\sigma(t)\,\tC+\tC\,\sigma^{\mathrm{t}}(t)
-\dfrac{1}{\tau_p}\left[f\big(\operatorname{tr}\tC\big)\tC
-R_0^2\mathcal{I}\right],
\end{equation}
where~$\mathcal{I}$ is the identity matrix and $\tC$ denotes the polymer 
conformation tensor calculated according to the Peterlin approximation.
If the flow is turbulent, both $\mathcal{C}$ and $\tC$ have a random behavior.
In the following, we shall denote:
\begin{equation}
\mathcal{T}\equiv f(\operatorname{tr}\mathcal{C})
\mathcal{C}
\qquad \text{and} \qquad
\tT\equiv f\big(\operatorname{tr}\tC\,\big)\tC.
\label{eq:T}
\end{equation}
Equation~\eqref{eq:FENE-P} describes the evolution of the conformation
tensor of a polymer along the Lagrangian trajectory of its center of mass.
Numerical simulations of drag reduction use the Eulerian counterpart
of Eq.~\eqref{eq:FENE-P}, which is obtained by replacing $d\tC/dt$
with $\partial_t\tC+\bm u\cdot\nabla\tC$ and $\sigma(t)$
with the Eulerian velocity gradient. In principle, the evolution 
equation for the conformation tensor
should be coupled with the Navier--Stokes equations through an additional
stress term proportional to $\nabla\cdot\tT$.~\cite{BHAC77}
Here, however, we focus
on the impact of the Peterlin approximation upon the statistics of polymer
deformation and consider passive polymers only (one-way coupling).
In the rest of the paper, we thus study 
the relation between Eq.~\eqref{eq:FENE-P} and
Eq.~\eqref{eq:dumbbell} when $\sigma(t)$ is given
by the incompressible Navier--Stokes equations in the turbulent regime.
Some considerations are useful to guide our study:
\begin{enumerate}
\item $\bm R(t)$ cannot be calculated from
the solution of the FENE-P equation
(Eq.~\eqref{eq:FENE-P}). By contrast,
$\mathcal{C}(t)$ can be constructed from the solution of the FENE equation
(Eq.~\eqref{eq:dumbbell}) by averaging the dyadic $R_i(t)R_j(t)$ over the 
realizations of the noise $\bm \xi(t)$. The tensor $\mathcal{C}(t)$ 
can then be compared with $\tC(t)$;
\item because of the
Peterlin approximation (Eq.~\eqref{eq:peterlin-closure}), the FENE and 
FENE-P equations yield a different evolution for the conformation tensor.
This holds for both laminar and turbulent flows;
\item
If the flow is turbulent, the statistics of $\bm R(t)$ differs from
that of $\mathcal{C}(t)$,
even if $\mathcal{C}(t)$ is calculated from Eq.~\eqref{eq:dumbbell} (and hence
no closure approximation is required).
Consider for example the random variables $R(t)$ and $\rho(t)\equiv
\sqrt{\langle R^2(t)\rangle_\xi}=
\sqrt{\operatorname{tr}\mathcal{C}(t)}$.
In general, the probability density function (PDF)
of $R(t)$ is different from that
of $\rho(t)$, as can be seen by noting that
$\langle\langle R^{\alpha}\rangle_\xi\rangle_\sigma
\neq\langle\langle R^2\rangle_\xi^{\frac{\alpha}{2}}\rangle_\sigma$
($\alpha\neq 2$), where
$\langle\cdot\rangle_\sigma$ denotes an average over the statistics of the
turbulent velocity gradient. 
\end{enumerate}
In conclusion,
the FENE and FENE-P models differ for two reasons:
the Peterlin approximation and the statistical effect due to the
fact that the statistics of $\bm R(t)$ cannot be deduced from that 
of $\mathcal{C}(t)$.
Hence, in the turbulent regime,
the proper way to examine the Peterlin approximation 
is to first construct $\mathcal{C}(t)$ from the FENE equation and
then compare its statistics with that of the solution of the FENE-P equation.
If the statistics of $\bm R(t)$ is directly compared with that of
$\tC(t)$, the error due to the Peterlin 
approximation is combined with the statistical effect discussed at point 3
above. This fact seems to have been overlooked in previous studies.

\section{Lagrangian simulations}
\label{sec:Lagrangian}

The dynamics of polymers is studied by using
a database of Lagrangian trajectories that was previously generated to
examine the dynamics of both tracer and inertial particles
in turbulent flows.~\cite{CCLT08,BBLST10}
The turbulent velocity field is obtained by
direct numerical simulation of the three-dimensional
incompressible Navier--Stokes equations:
\begin{equation}
\label{eq:NS}
\partial_t\bm u+\bm u\cdot\nabla\bm u=-\nabla p+\nu\nabla^2\bm u+\bm f,
\qquad \nabla\cdot\bm u=0,
\end{equation}
where $p$ is the pressure field and $\nu=2\times 10^{-3}$ 
is the kinematic viscosity. The forcing $\bm f$ is such that the 
spectral content of the first low-wavenumber shells remains constant 
in time. The domain is a three-dimensional periodic box of linear size
$2\pi$.
Equations~\eqref{eq:NS}
are solved by means of a fully dealiased pseudospectral
algorithm with second-order Adams--Bashforth time stepping.
The number of grid points is $512^3$, while the integration time step is 
$dt=4\times10^{-4}$.
In this simulation,
the Kolmogorov time is $\tau_\eta=4.7\times 10^{-2}$ and the Taylor-microscale Reynolds
number is $R_\lambda=185$
(for more details on the numerical simulation, see
Refs.~\onlinecite{CCLT08,BBLST10}). 
We expect our results not to depend
significantly on the value of $R_\lambda$ except for some residual effects induced by
intermittency in the statistics of the velocity gradients.~\cite{BBPVV91}

As mentioned in Sect.~\ref{sec:models}, the inertia of polymers is
negligible. Furthermore, their thermal diffusivity is very small
compared to the turbulent diffusivity.
Hence the center of mass of a polymer moves like a tracer
and its position $\bm x_c(t)$ satisfies the following equation:
\begin{equation}
\label{eq:center-of-mass}
\dfrac{d\bm x_c}{dt}=\bm u(\bm x_c(t),t).
\end{equation}
Equation~\eqref{eq:center-of-mass}
is once again solved by using a second-order Adams--Bashforth scheme;
a tri-linear interpolation algorithm is used to determine the value of
the velocity field at the position of the polymer.~\cite{CCLT08,BBLST10}
After the statistically stationary state is reached for both the fluid motion
and the translational dynamics of polymers, 
the positions of the center of mass of $128\times 10^3$
polymers are recorded every $\Delta t=10 dt=4\times 10^{-3}\approx
\tau_\eta/10$.
The total integration time after steady-state is $T=13.2$, 
which corresponds to 6 eddy turnover times approximately.

The velocity gradient $\sigma(t)$ is
recorded every $\Delta t$
along the trajectory of the center of mass of each polymer
and is inserted into Eqs.~\eqref{eq:dumbbell} and~\eqref{eq:FENE-P}
in order to determine the dynamics of the separation vector and of the 
conformation tensor.
Equation~\eqref{eq:dumbbell} is
solved by using the semi-implicit predictor--corrector method introduced
by \"Ottinger;~\cite{O96} the integration time step is equal to $\Delta t$
for all values of the parameters.
The initial condition for Eq.~\eqref{eq:dumbbell} is such that 
$R_i(0)=R_0/\sqrt{3}$, $i=1,2,3$.
Equation~\eqref{eq:FENE-P} for $\mathcal{C}^{\mathrm{P}}$ 
is integrated by means of the semi-implicit algorithm proposed in Ref.~\onlinecite{TDMS03},
which ensures that $\operatorname{tr}\tC\leqslant L^2$.

The Weissenberg number is defined
as~$\Wi\equiv\tau_p/\tau_\eta$ and is the ratio of the time
scales associated with the elastic force and with the velocity
gradient.  In our simulations, $\Wi$ varies between $10^{-2}$
and $10^2$. (An alternative definition of the
  Weissenberg number uses the maximum Lyapunov exponent of the flow,
  $\lambda$, to estimate the reciprocal of the stretching time
  associated with the velocity gradient.  In our simulations,
  $\lambda\approx 0.14\tau_\eta$.~\cite{BBBCMT06} Thus, the
  Weissenberg number based on the Lyapunov exponent is
  $\Wi_\lambda=\lambda\tau_p\approx 0.14 \Wi$).  The squares of the
equilibrium and maximum extensions of the polymer are $R_0^2=1$ and
$L^2=3\times 10^3$, as in Refs.~\onlinecite{JC07,WG10}.  The number of
realizations of thermal noise per Lagrangian trajectory is $2\times
10^3$.

Finally, the statistics of polymer deformation 
is collected over
the Lagrangian trajectories, over
the realizations of thermal noise, and over time (only for times greater than the
time required for $\bm R(t)$ to reach the statistically steady state).

We note that the statistics of the separation vector $\bm R$ in isotropic turbulence has been studied
thoroughly by Watanabe and Gotoh.~\cite{WG10}
The results on the statistics of $\bm R$ given below agree with those presented in Ref.~\onlinecite{WG10}.
Here, we compare the statistics of $\mathcal{C}$ with that of
$\mathcal{C}^{\mathrm{P}}$,
in order to determine the effect of the Peterlin approximation on the dynamics of polymers.

\begin{figure}[t]
\centering
\includegraphics[width=0.45\textwidth]{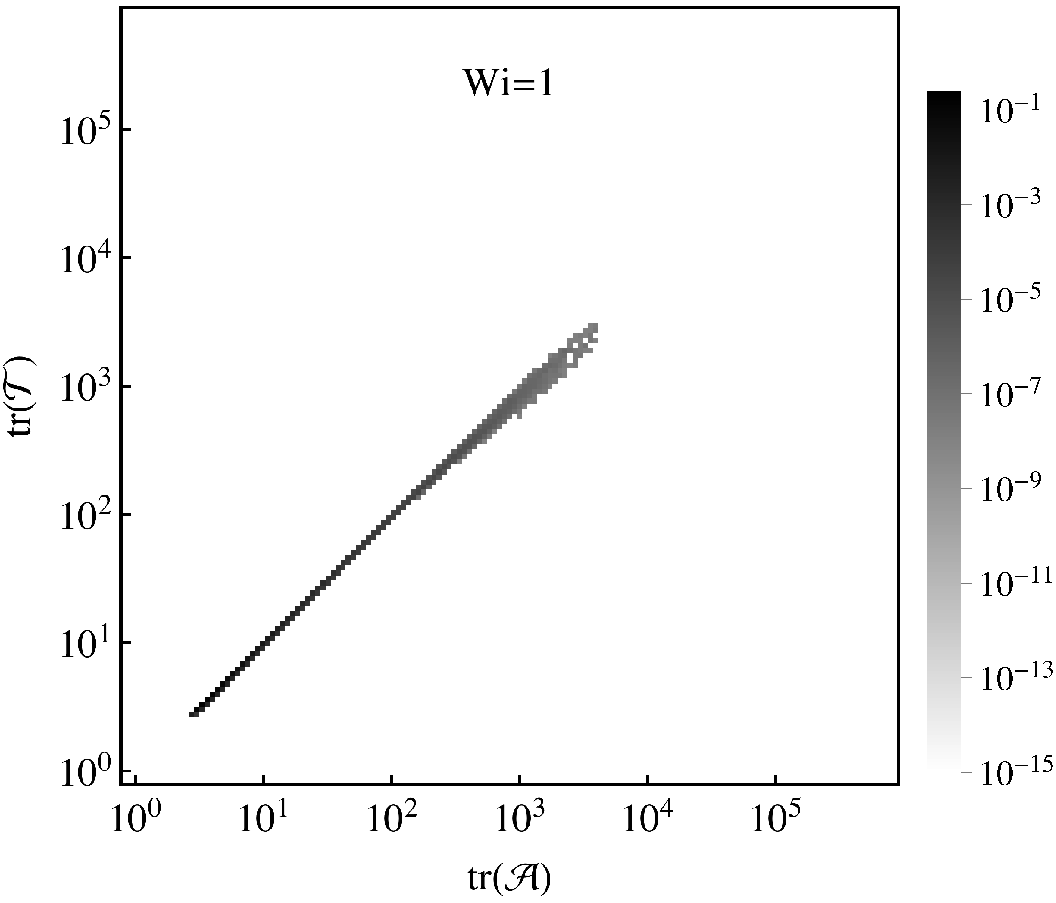}
\hspace{1cm}
\includegraphics[width=0.45\textwidth]{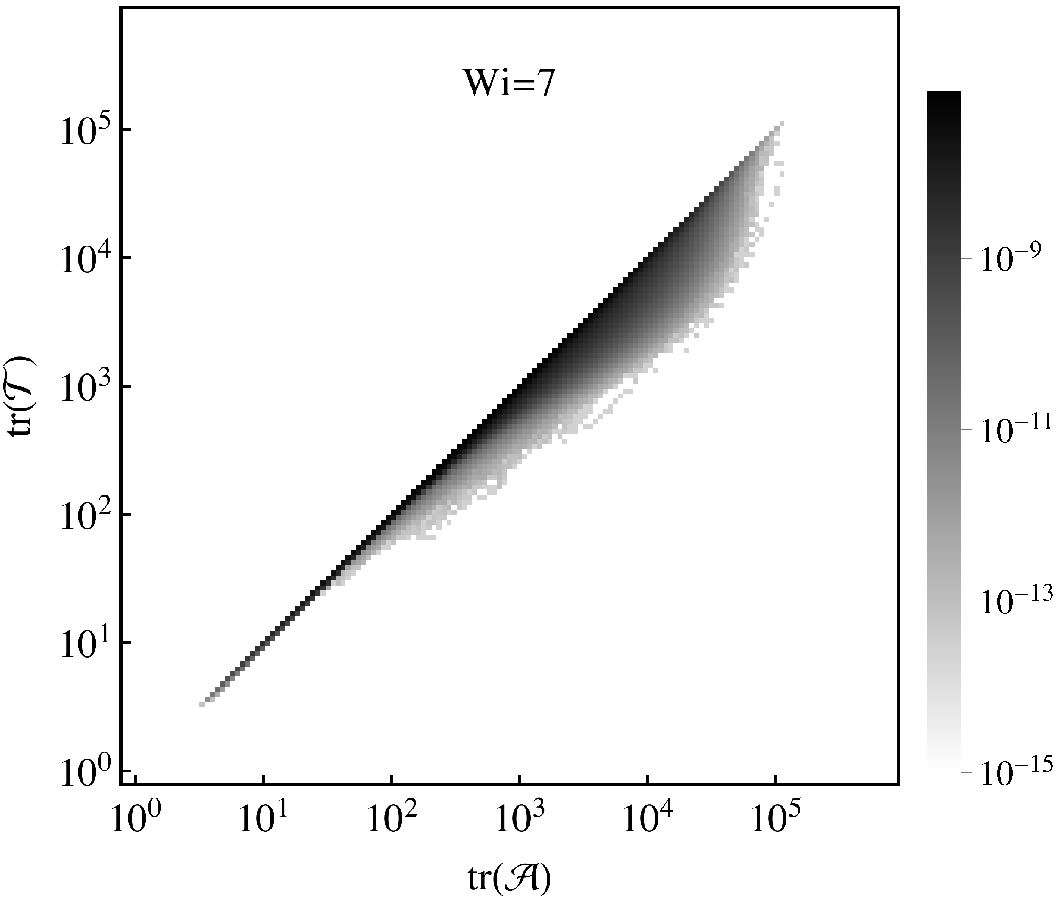}
\\
\begin{minipage}{0.45\textwidth}
\centering
\includegraphics[width=\textwidth]{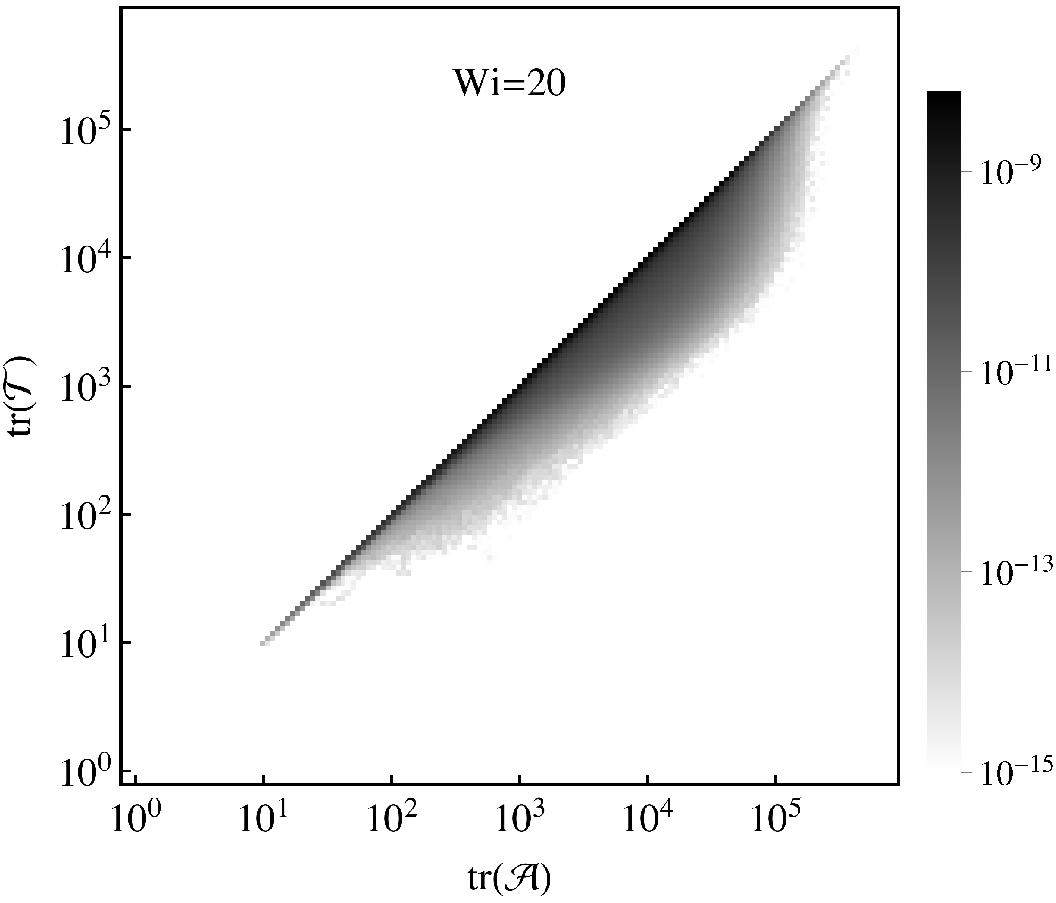}
\end{minipage}
\hspace{1cm}
\begin{minipage}{0.45\textwidth}
\includegraphics[width=0.9\textwidth]{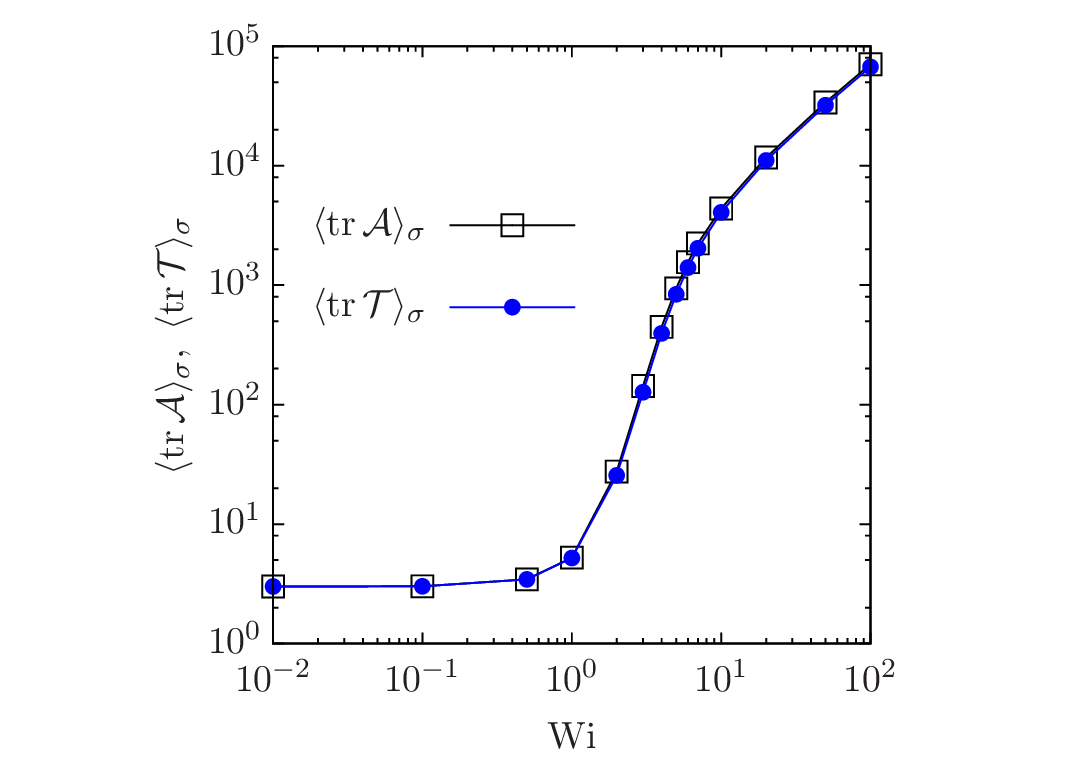}
\hfill
\end{minipage}
\caption{Contour plot of $P_\sigma(\operatorname{tr}\mathcal{A},\operatorname{tr}\mathcal{T})$ for $\Wi=1$ (top left panel),
$\Wi=7$ (top right panel) and $\Wi=20$ (bottom left panel). 
The bottom right panel
shows the averages of $\operatorname{tr}\mathcal{A}$ (black squares)
and $\operatorname{tr}\mathcal{T}$ (blue circles) 
as a function of the Weissenberg number.
}
\label{fig:joint}
\end{figure}

\section{Results}
\label{sec:results}

In this Section, we examine 
the statistics of polymer extension and orientation in the FENE and FENE-P models.
Before presenting the results, it is useful to define some notations.
If the statistics of a random variable depends both on thermal noise and
on the velocity gradient (as for instance in the case of $R$), its 
PDF is denoted as $P_{\xi,\sigma}$.
If the statistics of a random variable  (e.g., $\operatorname{tr}\mathcal{C}$)
only depends on the velocity
gradient, then its PDF is denoted
as $P_\sigma$.
The auto-correlation function of a scalar random variable $X(t)$ is 
denoted as $F_X(t)$ and the
correlation time of $X(t)$ is: $T_X=\int_0^\infty F_X(t)dt$.
The auto-correlation function of
a statistically isotropic, unit random vector  $\bm X(t)$ is defined as:
\begin{equation}
F_{\bm X}(t)=
2\langle \vert\bm X(t+t_0)\cdot\bm X(t_{0})\vert\rangle-1
\end{equation}
and the associated correlation time is defined as: $T_{\bm X}=\int_0^\infty F_{\bm X} (t)dt$.
We denote by $\hat{\bm e}_i$ ($i=1,2,3$) the unit eigenvectors of 
the rate-of-strain tensor $\mathcal{S}\equiv(\nabla\bm u+\nabla
\bm u^{\mathrm{T}})/2$; the unit eigenvectors $\hat{\bm e}_i$
are ordered by descending eigenvalue,
i.e. $\hat{\bm e}_1$ is associated with the largest eigenvalue of $\mathcal{S}$
and $\hat{\bm e}_3$ with the smallest one.
The direction of vorticity is $\hat{\bm\omega}$.
Finally, we denote by $\hat{\bm z}_1$ and $\hat{\bm z}^{\mathrm{P}}_1$ the first 
unit eigenvector of $\mathcal{C}$ and $\mathcal{C}^{\mathrm{P}}$, respectively.

\subsection{The Peterlin approximation}

As mentioned in Sect.~\ref{sec:models}, the Peterlin approximation 
consists in replacing $\mathcal{A}$ with $\mathcal{T}$ in the evolution equation
for the polymer conformation tensor ($\mathcal{A}$ 
and $\mathcal{T}$ have been defined in Eqs.~\eqref{eq:peterlin-closure} 
and~\eqref{eq:T} and, up to a concentration-dependent 
multiplicative constant, 
are the polymer stress tensors in the FENE and in the FENE-P models, 
respectively). 
The eigenvectors of $\mathcal{A}$ and $\mathcal{T}$ are the same, but their
eigenvalues may differ. Indeed, Jensen's inequality yields:
$\operatorname{tr}\mathcal{A}\geqslant\operatorname{tr}\mathcal{T}$.
A first indication of the effect of the Peterlin 
approximation is thus given by the joint PDF
$P_\sigma(\operatorname{tr}\mathcal{A},\operatorname{tr}\mathcal{T})$
(see Fig.~\ref{fig:joint}).
For small polymer extensions or for small values of $\Wi$,
$\operatorname{tr}\mathcal{A}$ and 
$\operatorname{tr}\mathcal{T}$ are approximately the same because 
$1-R^2/L^2\approx 1$; hence the Peterlin approximation holds very well.
By contrast, for large extensions or for large values of $\Wi$, 
the deviations
of $\operatorname{tr}\mathcal{T}$ from $\operatorname{tr}\mathcal{A}$ are significant, 
i.e. the Peterlin approximation is poor.
Notwithstanding, 
$\langle\operatorname{tr}\mathcal{T}\rangle_\sigma$ 
and $\langle\operatorname{tr}\mathcal{A}\rangle_\sigma$ do not differ appreciably
(see the bottom, right panel in Fig.~\ref{fig:joint}).
This fact demonstrates that the study of average values may not
suffice to investigate the validity of the Peterlin approximation.
The full statistics of the separation vector and the conformation tensor
should be investigated, which requires following the Lagrangian dynamics of a large
number of polymers.

\begin{figure}[t]
\centering
\includegraphics[width=0.49\textwidth]{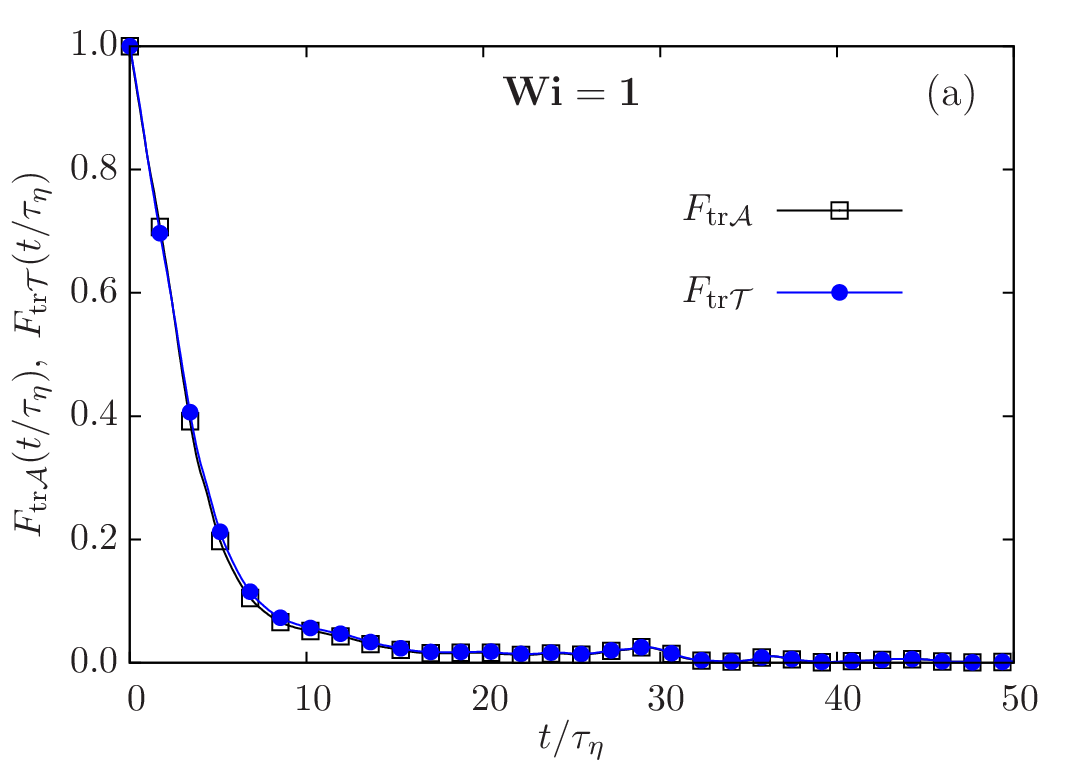}%
\hfill%
\includegraphics[width=0.49\textwidth]{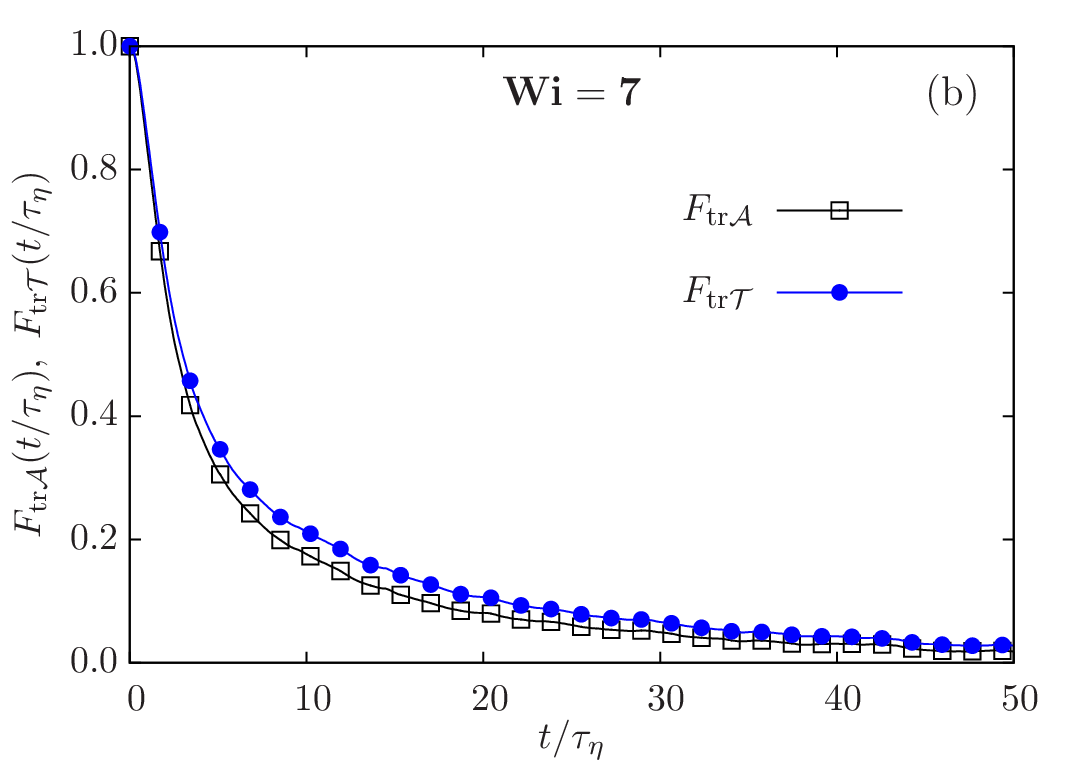}%
\\
\includegraphics[width=0.49\textwidth]{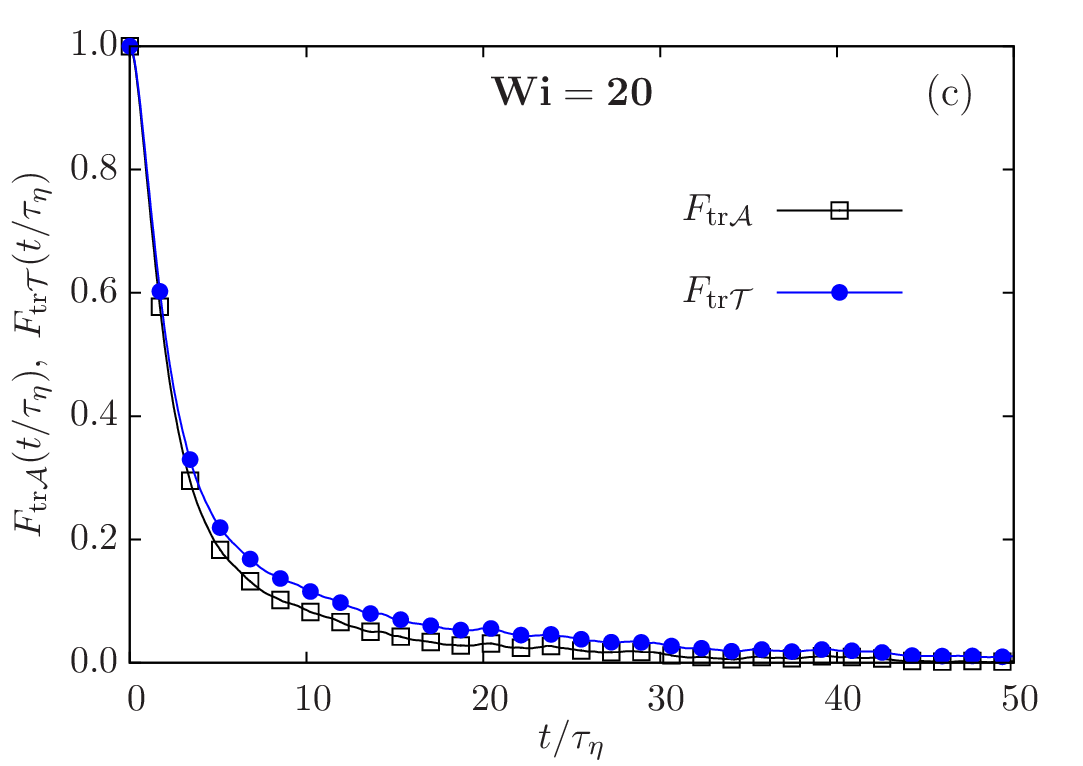}%
\hfill%
\includegraphics[width=0.49\textwidth]{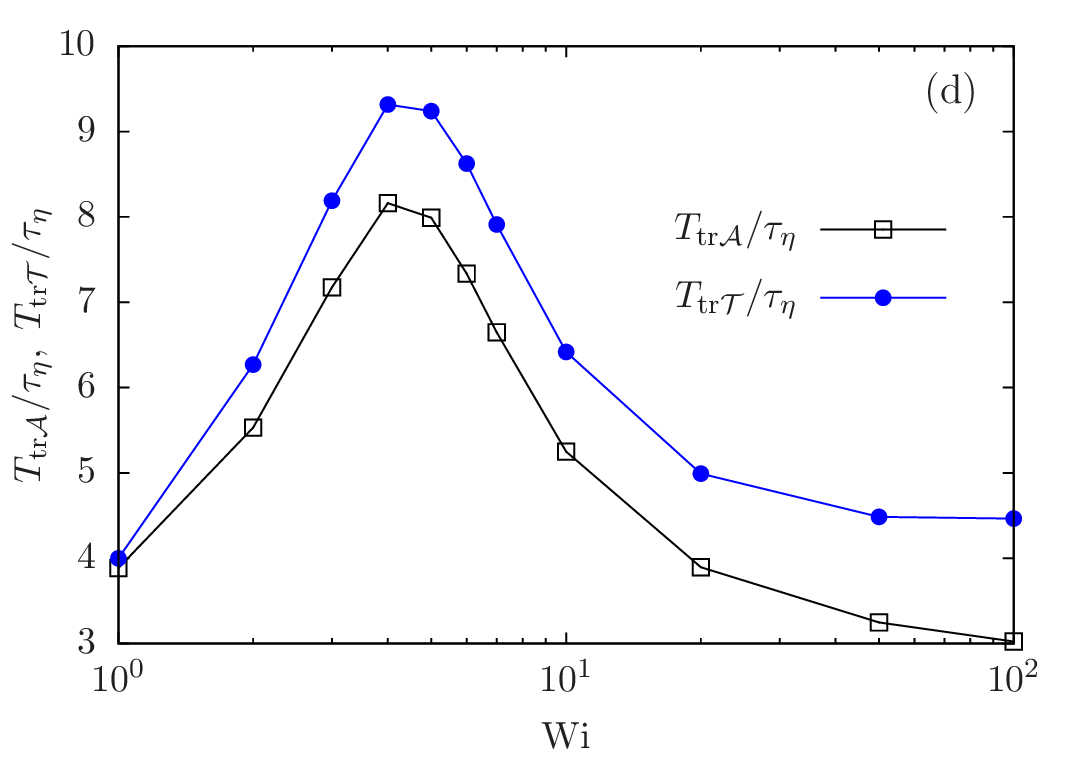}%
\caption{Autocorrelation function of $\operatorname{tr}\mathcal{A}$ (black squares) and
$\operatorname{tr}\mathcal{T}$ (blue circles).
for (a) $\Wi=1$, (b) $\Wi=7$, (c) $\Wi=20$.
Panel (d) shows the correlation times $T_{\operatorname{tr}\mathcal{A}}$ (black squares) and
$T_{\operatorname{tr}\mathcal{T}}$ (blue circles) rescaled by $\tau_\eta$ as a function of $\Wi$.}
\label{fig:corr-trace-nlin}
\end{figure}

In addition, the qualitative behavior of the temporal autocorrelations of $\operatorname{tr}\mathcal{A}$
and $\operatorname{tr}\mathcal{T}$ are similar
(Figs.~\ref{fig:corr-trace-nlin}(a) to~\ref{fig:corr-trace-nlin}(c)),
but for large $\Wi$ the correlation time
of $\operatorname{tr}\mathcal{A}$ is shorter than that of $\operatorname{tr}\mathcal{T}$ (Fig.~\ref{fig:corr-trace-nlin}(d)).

\subsection{Statistics of polymer extension}

\begin{figure}[t!]
\centering
\includegraphics[width=0.49\textwidth]{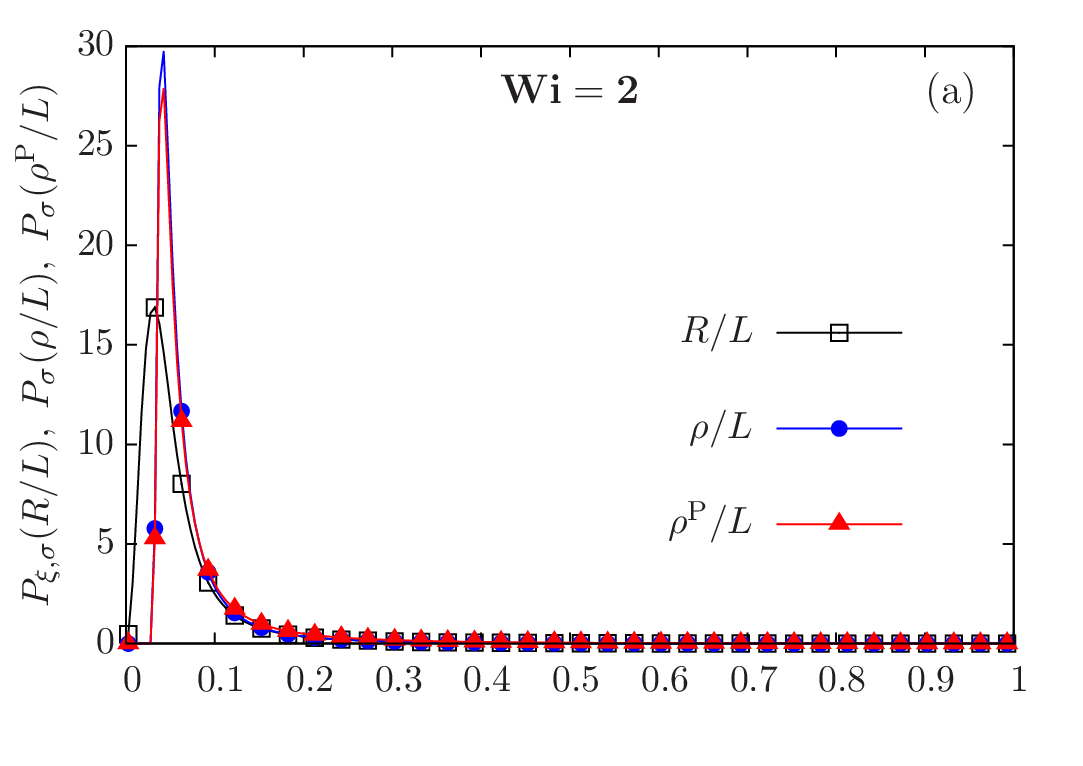}%
\hfill%
\includegraphics[width=0.49\textwidth]{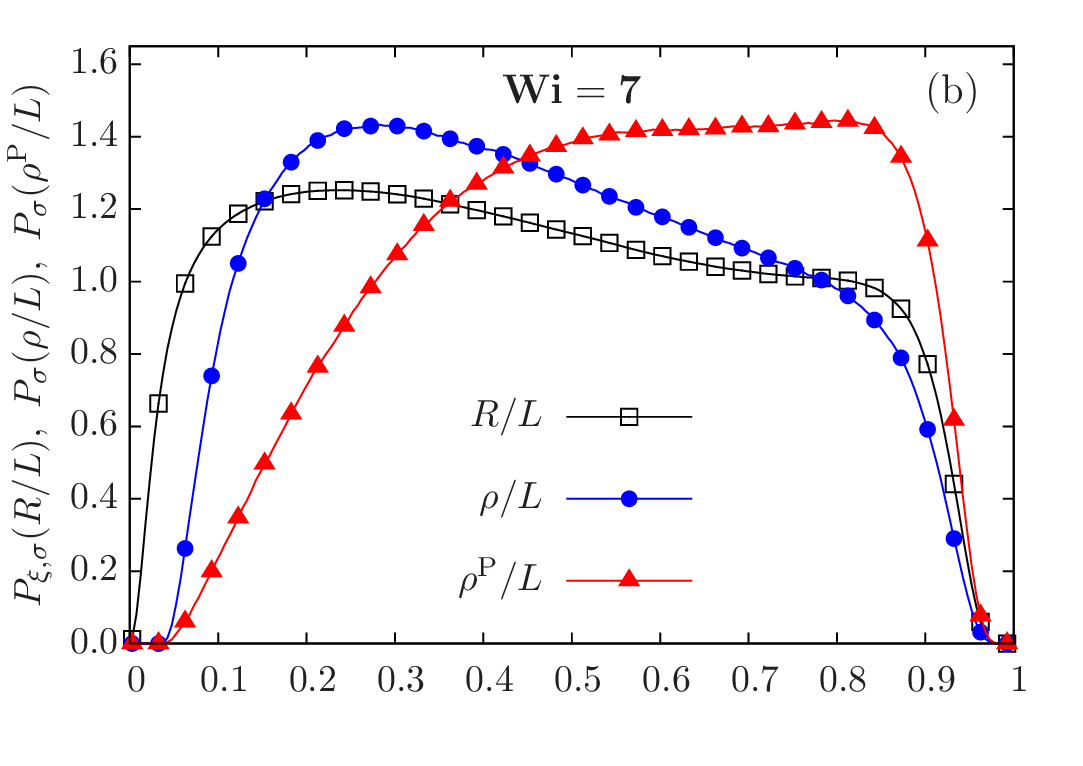}%
\\
\includegraphics[width=0.49\textwidth]{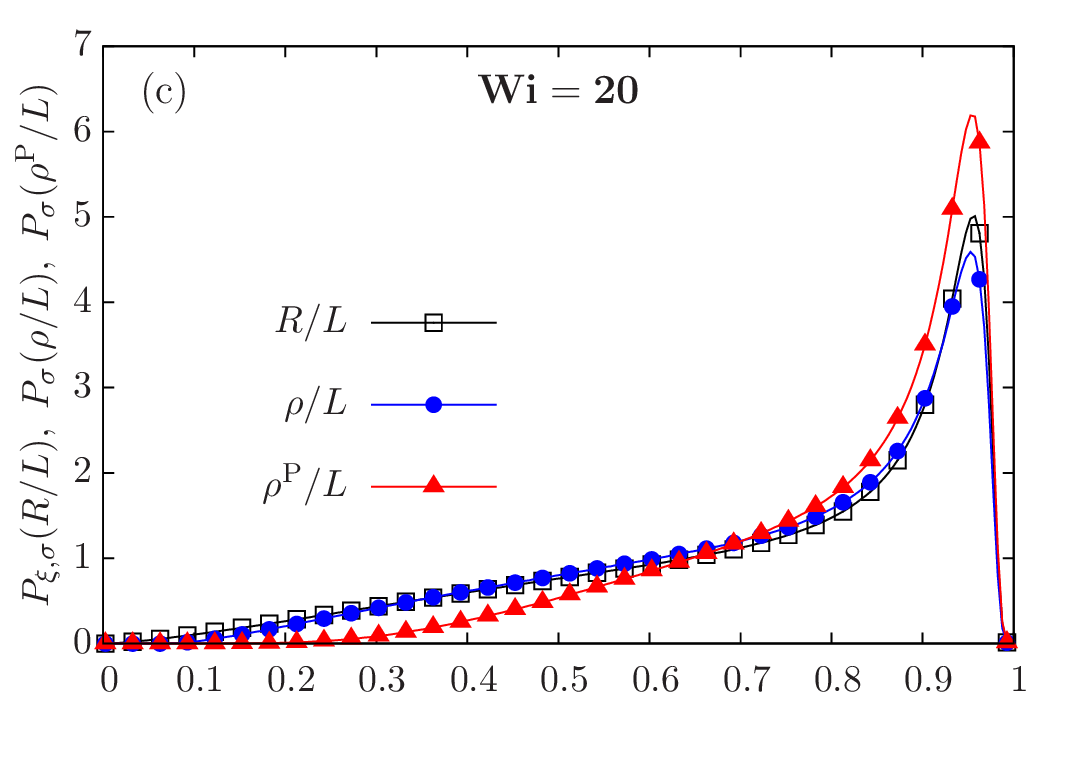}%
\hfill%
\includegraphics[width=0.49\textwidth]{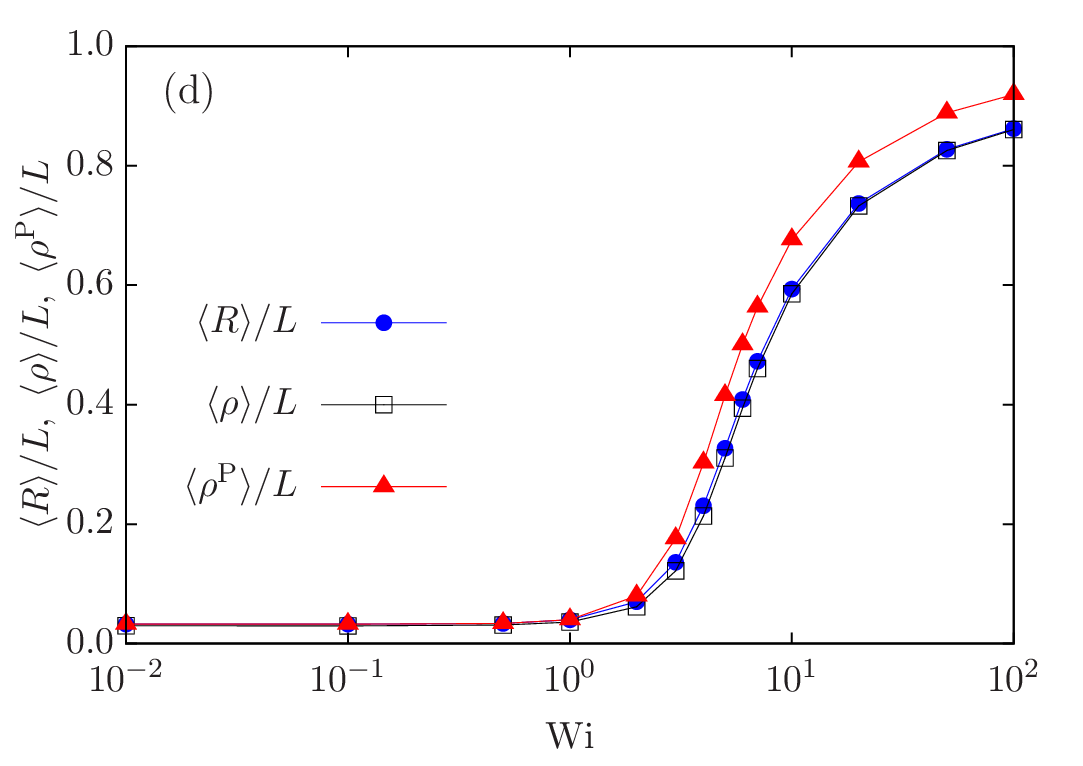}%
\caption{PDF of $R/L$ (black squares), $\rho/L$ (blue circles),
$\rho^\p/L$ (red triangles) for (a) $\Wi=2$, (b) $\Wi=7$,
(c) $\Wi=20$.
Panel (d) shows $\langle R\rangle_{\xi,\sigma}/L$ (black squares),
$\langle\rho\rangle_\sigma/L$ (blue circles),
$\langle\rho^\p\rangle_\sigma/L$ (red triangles) as a function of the Weissenberg number.}
\label{fig:pdf-extension}
\end{figure}
\begin{figure}[h]
\centering
\includegraphics[width=0.33\textwidth]{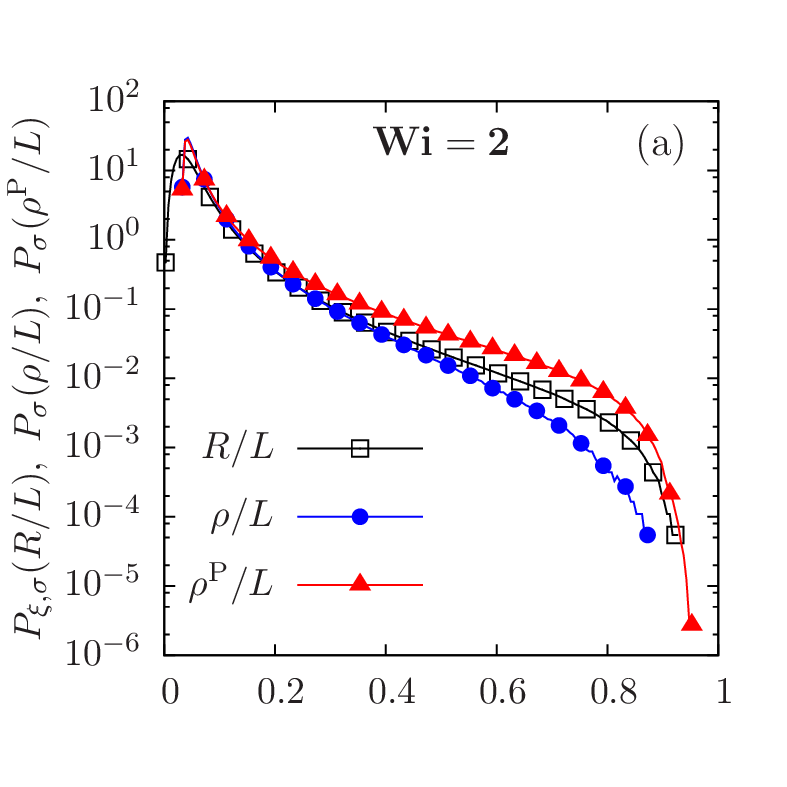}%
\hfill%
\psfrag{y}[b]{}%
\includegraphics[width=0.33\textwidth]{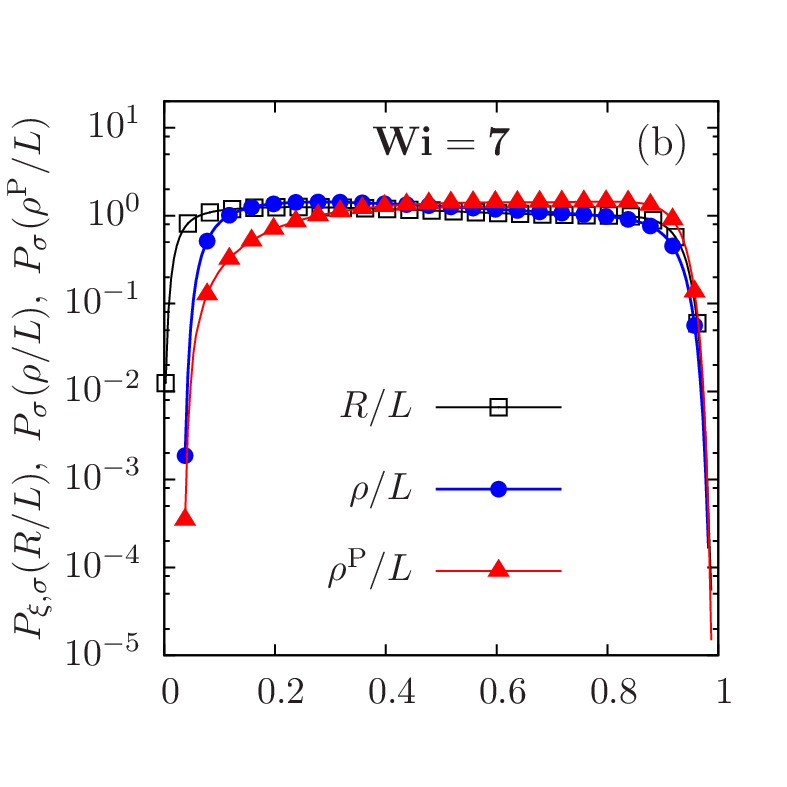}%
\hfill%
\psfrag{y}[b]{}%
\psfrag{a}[t]{$\Wi$}%
\includegraphics[width=0.33\textwidth]{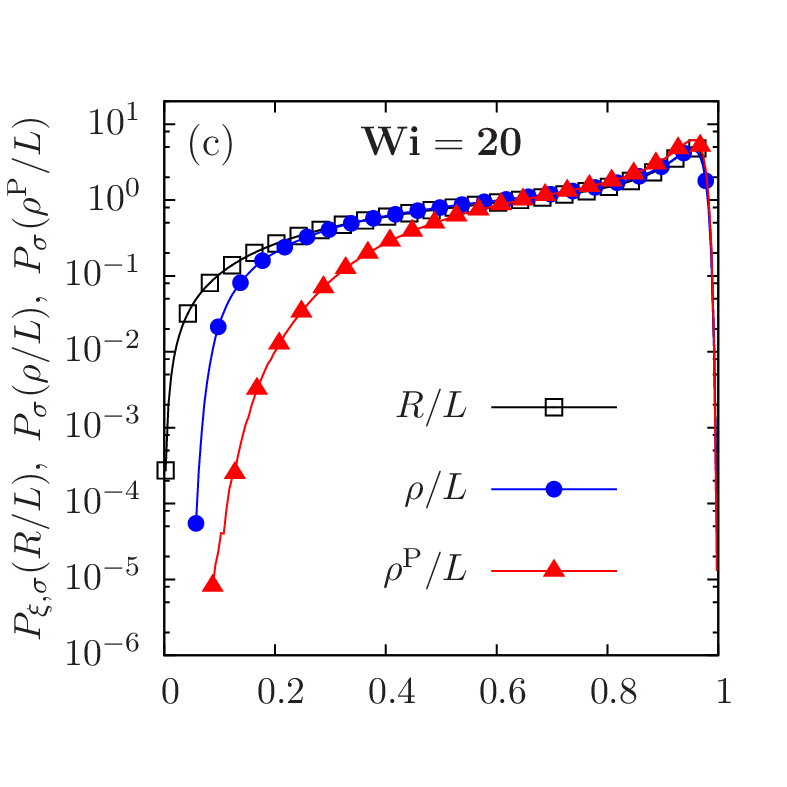}%
\caption{PDF of $R/L$ (black squares), $\rho/L$ (blue circles),
$\rho^\p/L$ (red triangles) in semi-logarithmic scale
for (a) $\Wi=2$, (b) $\Wi=7$, (c) $\Wi=20$.}
\label{fig:pdf-extension-log}
\end{figure}

The statistical properties of the separation $R$ are well understood.
For small values of $\Wi$, most of polymers are in the coiled state,
i.e. their extension is close to the equilibrium one.
Accordingly, the PDF of $R$ has a pronounced peak at $R_0$.
As $\Wi$ increases, polymers unravel and become more and more extended.
The transition from the coiled to the stretched state occurs when the Lyapunov
exponent of the flow exceeds $1/2\tau_p$, i.e. at $\Wi_\lambda=1/2$.~\cite{BFL00} 
(Note that some authors
define the Weissenberg number in terms of the
time scale associated with the exponential relaxation of $\sqrt{\langle
R^2(t)\rangle}$ instead of $\langle R^2(t)\rangle$ and hence obtain
a critical Weissenberg number equal to unity.)

At intermediate extensions, the PDF of $R$ has
a power-law behavior, i.e. $P_{\xi,\sigma}(R)\sim R^{-1-\alpha}$ for $R_0\ll R\ll L$.~\cite{BFL00}
This property of $P_{\xi,\sigma}(R)$ indicates that polymers with very different
extensions coexist in the fluid; whether the coiled or the stretched state dominates
depends on the value of $\Wi$.
The exponent $\alpha$ is positive in the coiled state and
decreases as a function of $\Wi$.~\cite{BFL00,WG10,MV11,MAV05}
As long as $\alpha>0$, the FENE dumbbell model reaches a steady-state even as $L$ tends to infinity
(the $L\to\infty$ limit of the FENE model is known as the Hookean model~\cite{BHAC77}).
However, when $\alpha$ vanishes a steady-state
PDF of the extension no longer exists if $L\to\infty$.
This behavior is interpreted as the coil--stretch transition.~\cite{BFL00}
Finally, if $\Wi$ increases beyond the value of the coil--stretch transition,
$\alpha$ becomes negative and
the maximum of $P_{\xi,\sigma}(R)$ moves from close to $R_0$ to close to $L$.\cite{MAV05,WG10}
The statistics of $R$ is shown in Figs.~\ref{fig:pdf-extension}
and~\ref{fig:pdf-extension-log} for different values of $\Wi$.

We noted in Sect.~\ref{sec:models} that
the comparison between the FENE and the FENE-P model
ought to be done in terms of the conformation tensors $\C$ and $\tC$
(rather than in terms of $R$ and $\tC$). 
Let us denote $\rho(t)=\sqrt{\operatorname{tr}\mathcal{C}(t)}$ and
$\rho^{\mathrm{P}}(t)=\sqrt{\operatorname{tr}\tC(t)}$.
To examine the influence of the Peterlin approximation on the statistics
of polymer extension, we calculate $\rho$
from the solution of Eq.~\eqref{eq:dumbbell} and $\rho^\p$ from Eq.~\eqref{eq:FENE-P}.
We then compare $P_\sigma(\rho/L)$ and $P_\sigma(\rho^{\mathrm{P}}/L)$ in the steady state
for different values of $\Wi$.
The plots shown in Figs.~\ref{fig:pdf-extension} and~\ref{fig:pdf-extension-log}
correspond to the coiled state ($\Wi=2$), 
the coil--stretch transition ($\Wi=7$), and the stretched state ($\Wi=20$).

\begin{figure}[t!]
\centering
\psfrag{x}[t]{$t/\tau_\eta$}%
\psfrag{a}[cr]{$F_R(t)$}%
\psfrag{b}[cr]{$F_\rho(t)$}%
\psfrag{c}[cr]{$F_{\rho^\p}(t)$}%
\includegraphics[width=0.49\textwidth]{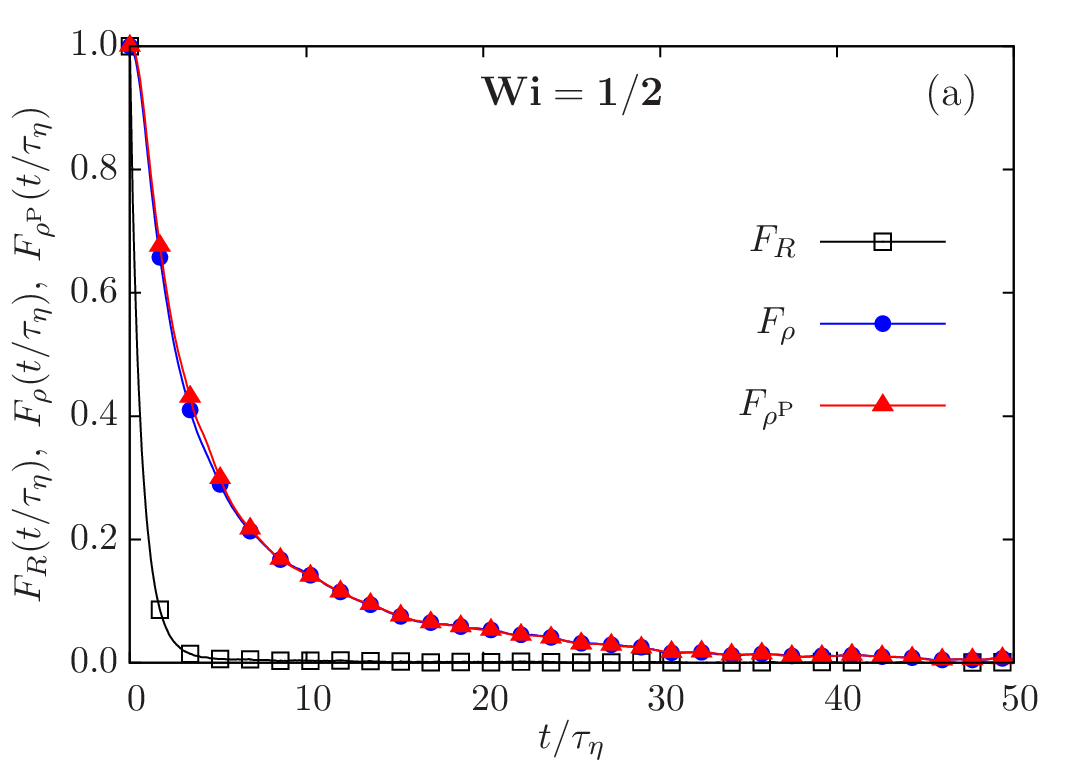}%
\hfill%
\includegraphics[width=0.49\textwidth]{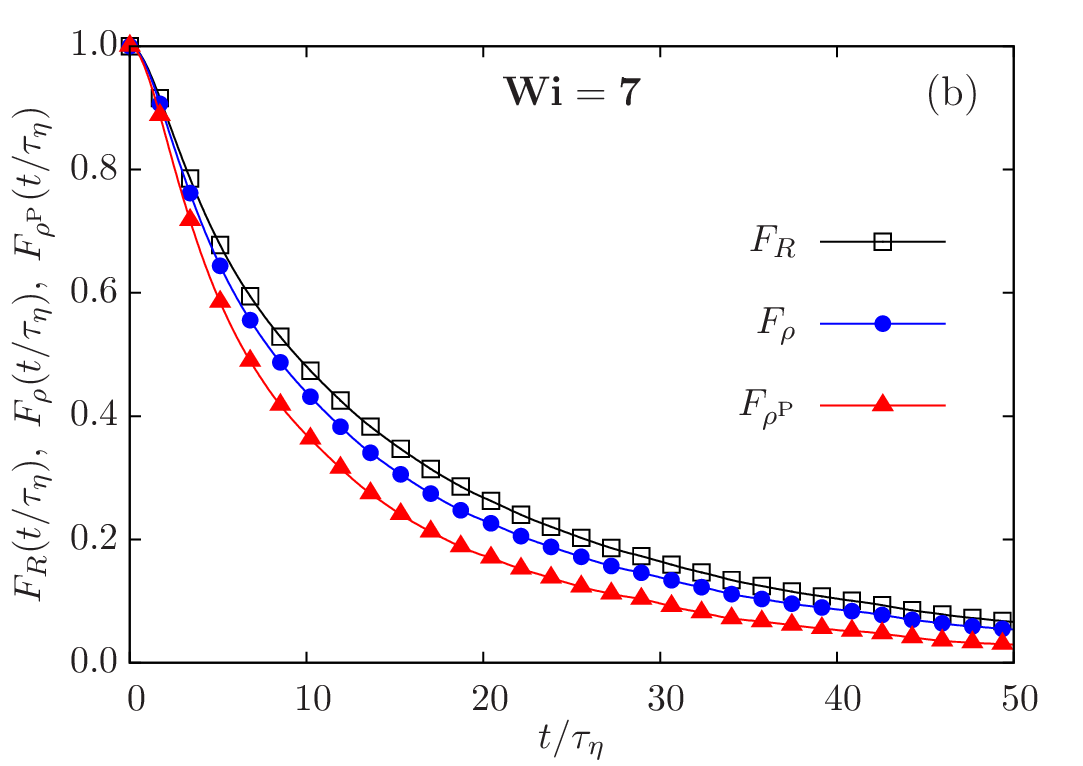}%
\\
\includegraphics[width=0.49\textwidth]{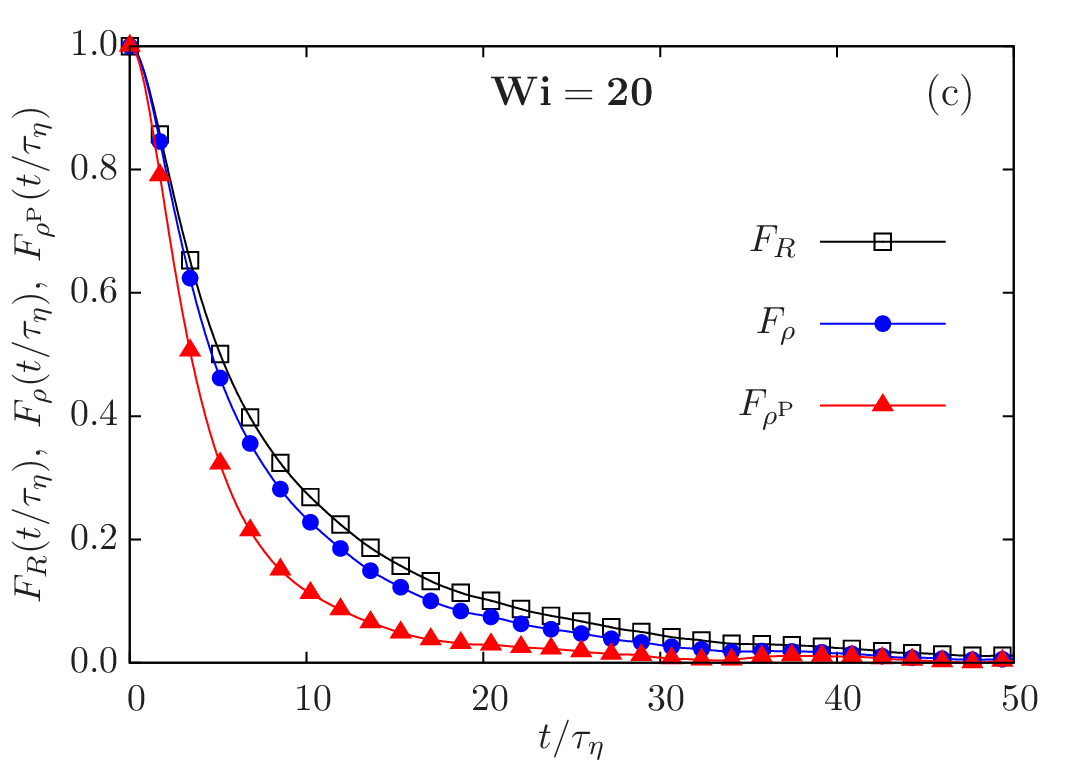}%
\hfill%
\includegraphics[width=0.49\textwidth]{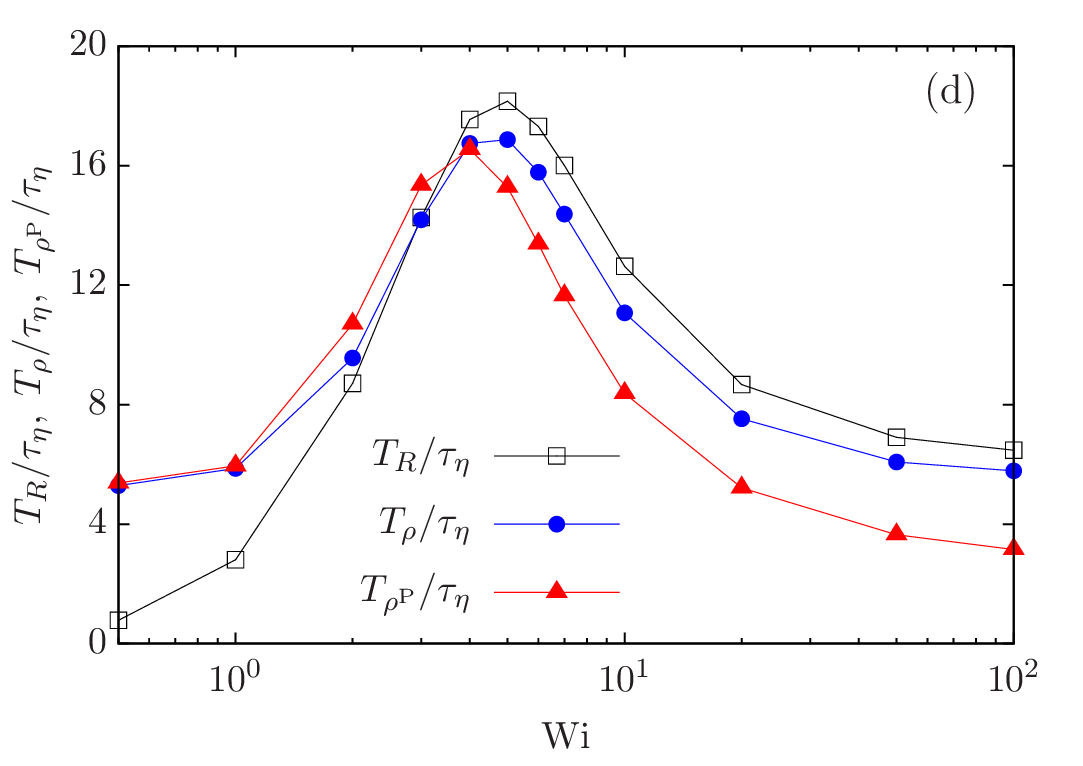}%
\caption{Autocorrelation function of $R$ (black squares), 
$\rho$ (blue circles),
$\rho^\p$ (red triangles) for (a) $\Wi=0.5$, (b) $\Wi=7$,
(c) $\Wi=20$.
Panel (d) shows the correlation times $T_R$ (black squares),
$T_\rho$ (blue circles), $T_{\rho^\p}$ (red triangles) rescaled by $\tau_\eta$
as a function of the Weissenberg number.}
\label{fig:corr-extension}
\end{figure}

For small values of $\Wi$, $P_\sigma(\rho/L)$ and $P_\sigma(\rho^{\mathrm{P}}/L)$ 
do not differ significantly, because the extension of most of polymers is
near to $R_0$ and hence  $1-R^2/L\approx 1$ and $\mathcal{C}\approx\mathcal{C}^\p$ (Fig.~\ref{fig:pdf-extension}(a)).
For intermediate and large values of $\Wi$, the statistics of $R$ is characterized by a broad distribution of
extensions around the mean value. The differences between 
$P_\sigma(\rho/L)$ and $P_\sigma(\rho^{\mathrm{P}}/L)$, therefore, are considerable.
In particular, since $\langle R^2\rangle_{\xi,\sigma}$ is appreciably less than $L^2$ even
for the largest $\Wi$ and $f(\zeta)$ rapidly diverges for $\zeta$ close to $L$ [see Eq.~\eqref{eq:elastic-force}], 
the restoring term in 
Eq.~\eqref{eq:FENE-P} is weaker than that in Eq.~\eqref{eq:dumbbell}.
Hence, in the FENE-P model large extensions are more probable than in the FENE model
to the detriment of small and intermediate extensions (Fig.~\ref{fig:pdf-extension}(b) and (c)).  
As a consequence, the 
FENE-P model overestimates the average extension of polymers for large values
of $\Wi$ (Fig.~\ref{fig:pdf-extension}(d)).
An analogous behavior of the average extension has been observed in
turbulent channel flows.~\cite{TDMS03,GSK04}

We also note that whereas for large $\Wi$ the PDFs of $R$ and $\rho$ are approximately the same 
(Fig.~\ref{fig:pdf-extension}(c)),
they are significantly different for intermediate or small $\Wi$ (Figs.~\ref{fig:pdf-extension}(a) and (b)). 
Indeed, in the former case, the stretching action of the velocity gradient is
very strong compared to the effect of thermal fluctuations,
and in most realizations $R\approx\sqrt{\tr\C}$.
In the latter case, the effect of thermal
fluctuations cannot be disregarded and the differences in the statistics
of $R$ and $\rho$ (see Sect.~\ref{sec:models}) become evident.
Furthermore, $P_{\xi, \sigma}(R/L)$ and $P_\sigma(\rho/L)$ mainly differ
for small and intermediate extensions,
because the large extensions are obtained in those realizations in which the velocity gradient 
is very intense and thermal noise can be neglected.
The above results demonstrate that comparing the statistics of $\bm R$ directly with that of $\tC$
may lead to wrong conclusions; indeed, at small $\Wi$,
$P_{\xi,\sigma}(R/L)$ and $P_\sigma(\rho^\p/L)$ are clearly different, whereas
$P_\sigma(\rho/L)$ and $P_\sigma(\rho^\p/L)$ are close.

The autocorrelation function of the extension is approximately exponential
both in the FENE and in the FENE-P model. 
However, $F_{\rho^\p}(t)$ is
a good approximation of $F_\rho(t)$ only for small $\Wi$ 
(Fig.~\ref{fig:corr-extension}(a)).
In addition, the FENE-P model captures the critical slowing down of 
polymers near the coil--stretch transition,~\cite{CPV06,VB06,GS08} but for
large $\Wi$ it underestimates the 
correlation time of the extension (Fig.~\ref{fig:corr-extension}(d)). Once again, we note that, for small
$\Wi$, a direct
comparison between $F_{R}(t)$ and $F_{\rho^\p}(t)$ would lead to
wrong conclusions about 
the effect of the Peterlin approximation on the temporal
statistics of polymer extension.

\subsection{Statistics of polymer orientation}

\begin{figure}[t!]
\centering
\includegraphics[width=0.49\textwidth]{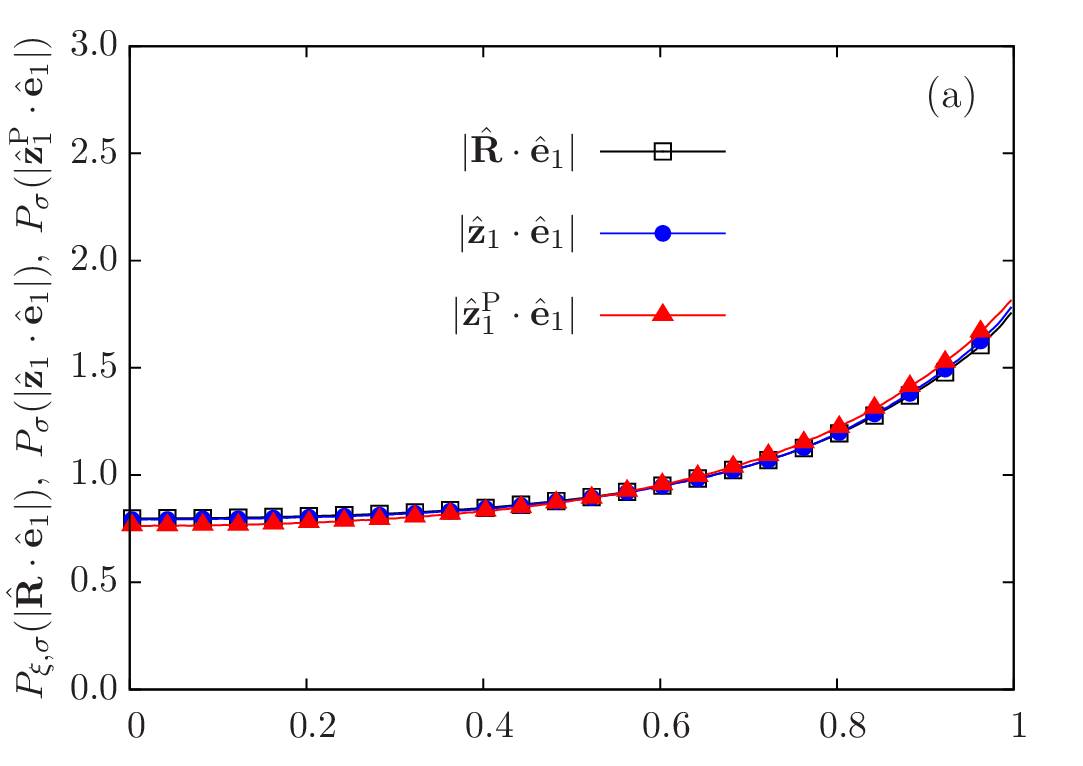}%
\hfill%
\includegraphics[width=0.49\textwidth]{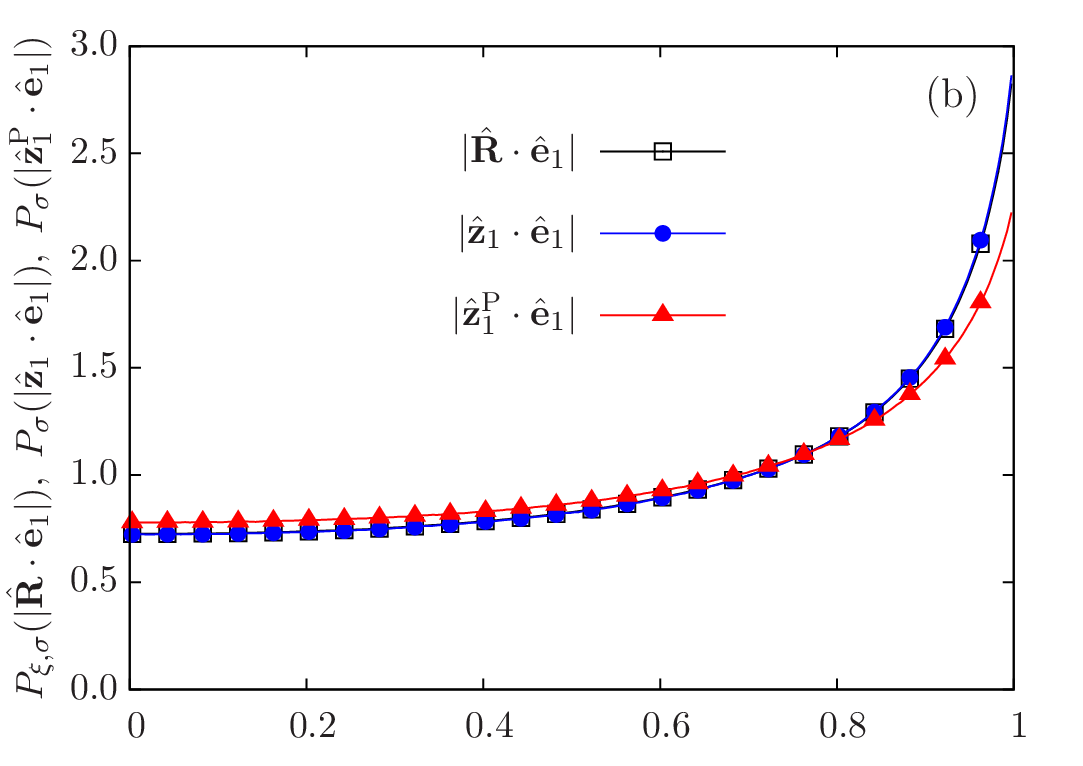}%
\\[2mm]
\includegraphics[width=0.49\textwidth]{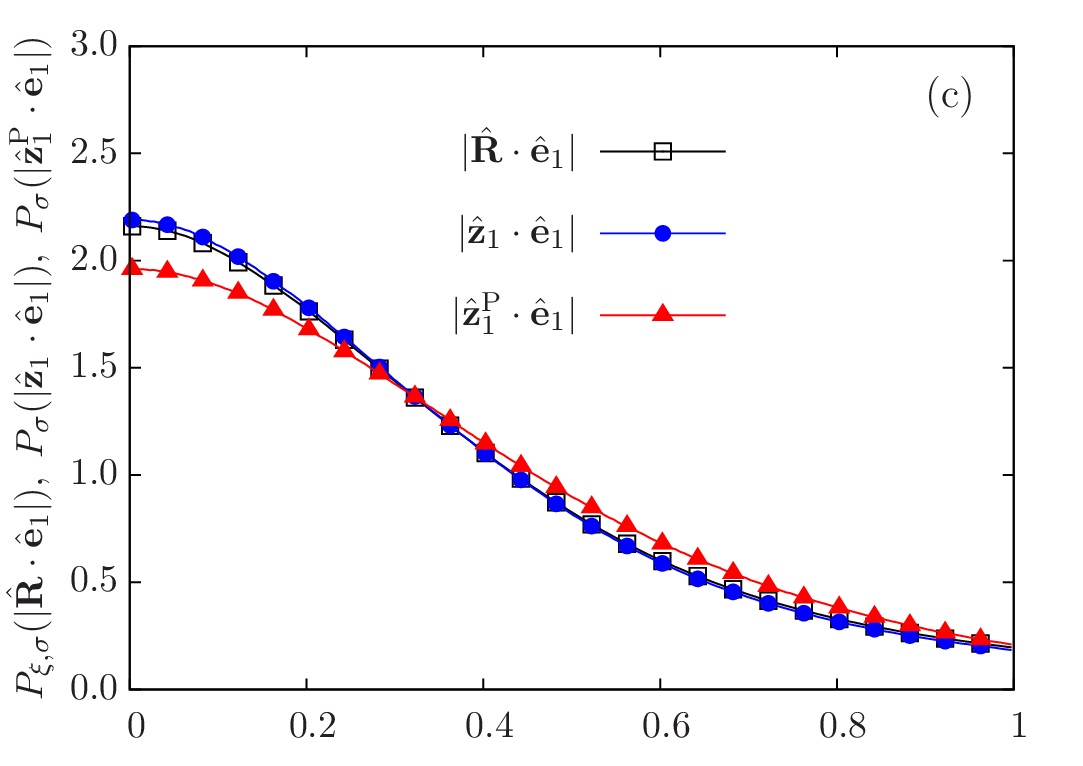}%
\hfill%
\includegraphics[width=0.49\textwidth]{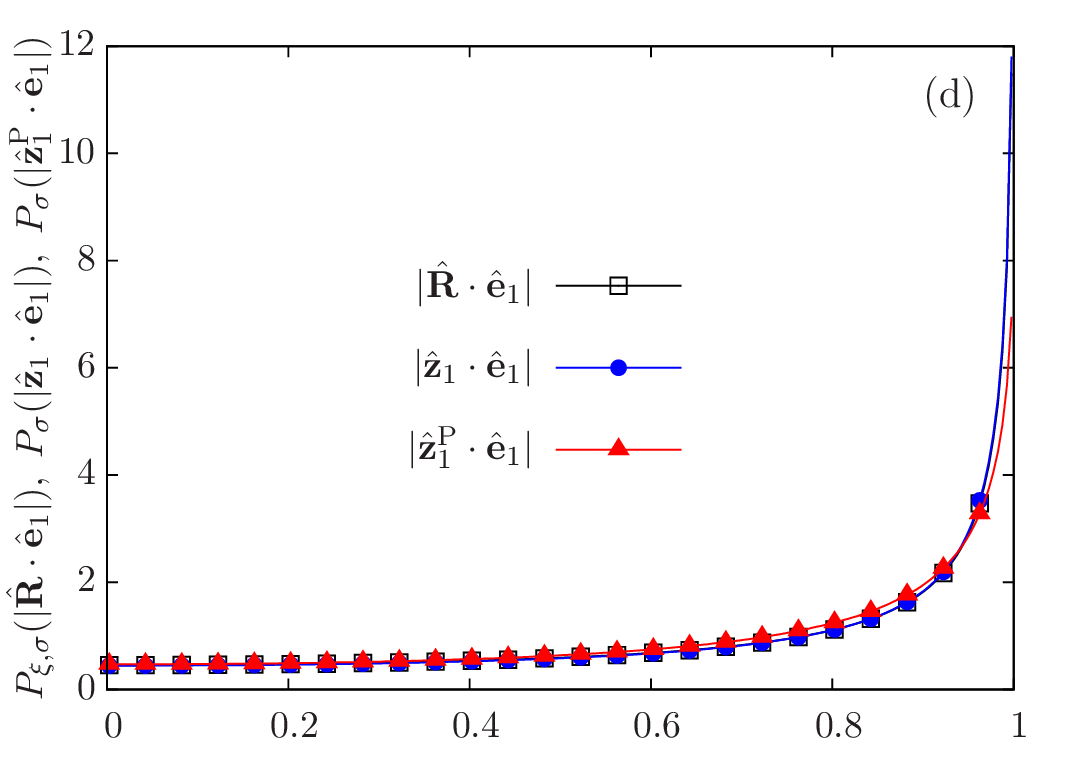}%
\caption{PDF of alignment with (a) $\hat{\bm e}_1$, 
(b) $\hat{\bm e}_2$, (c) $\hat{\bm e}_3$ and (d) $\hat{\bm \omega}$
for $\Wi=20$.} 
\label{fig:pdf-orientation}
\end{figure}
\begin{figure}[t!]
\centering
\includegraphics[width=0.49\textwidth]{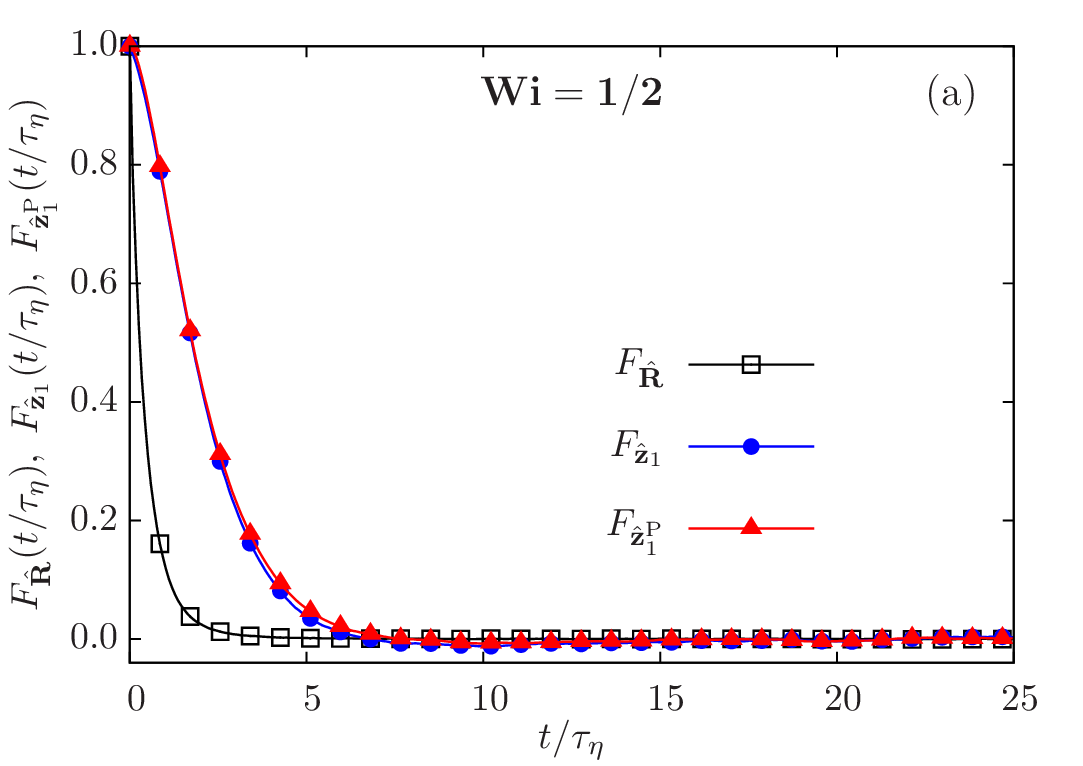}%
\hfill%
\includegraphics[width=0.49\textwidth]{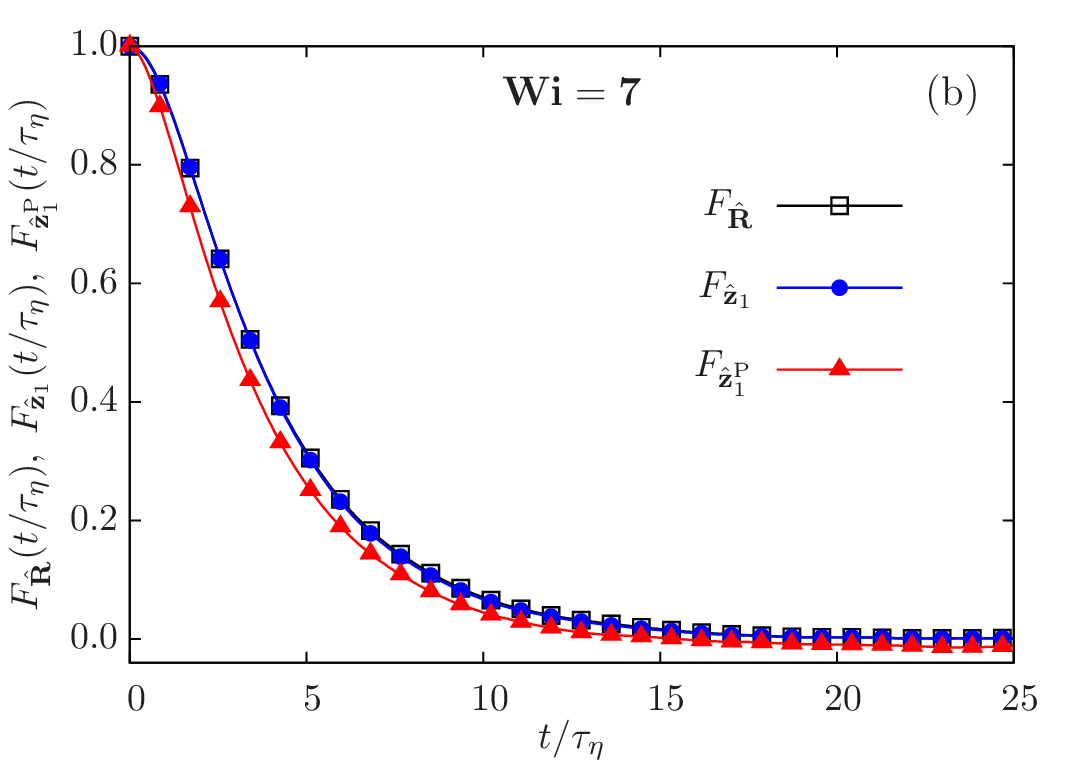}%
\\
\includegraphics[width=0.49\textwidth]{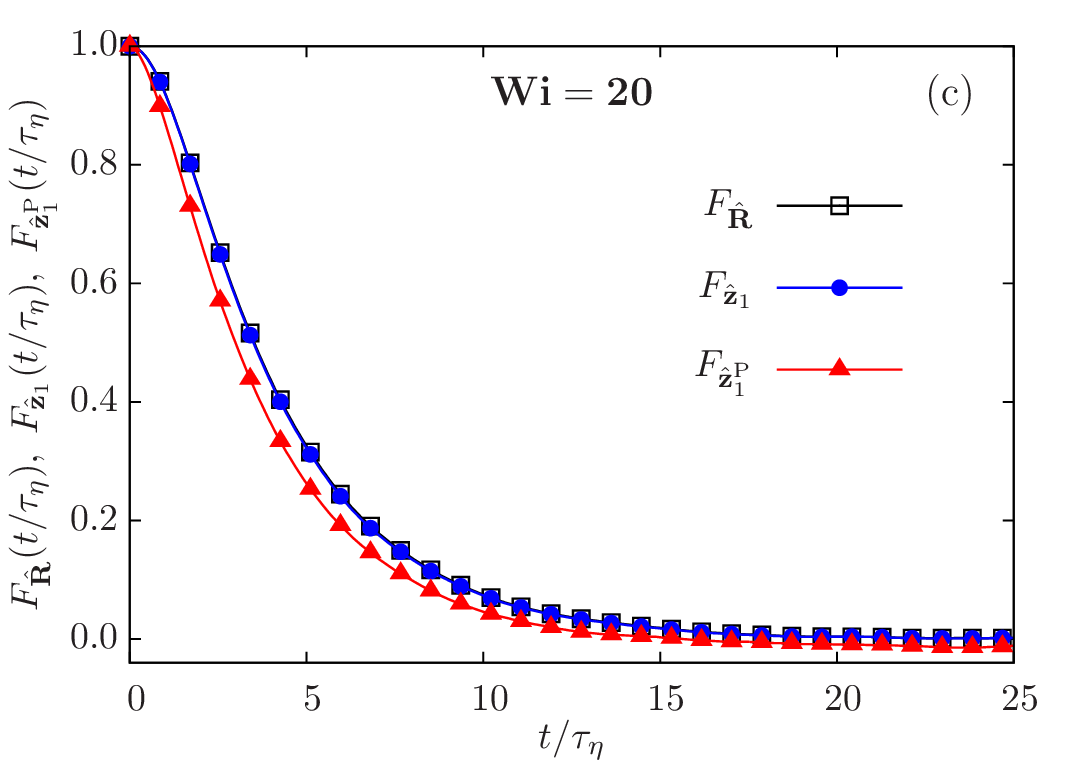}%
\hfill%
\includegraphics[width=0.49\textwidth]{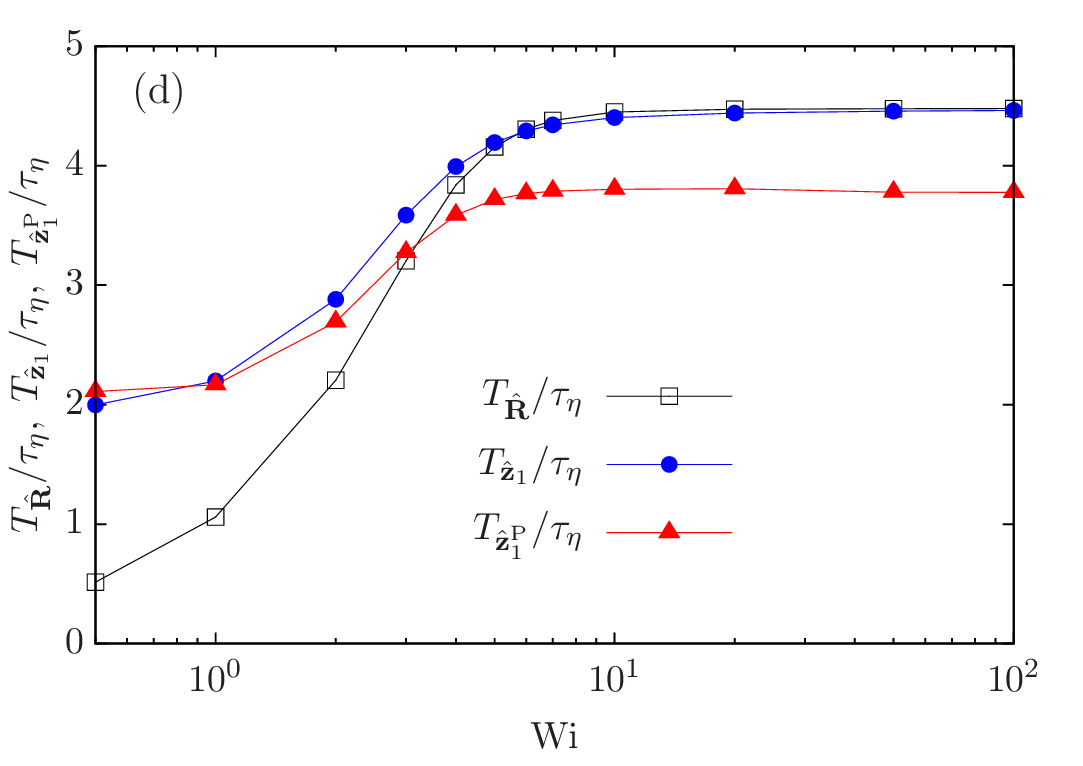}%
\caption{Autocorrelation function of $\hat{\bm R}$ (black squares), 
$\hat{\bm z}_1$ (blue circles),
$\hat{\bm z}_1^\p$ (red triangles) for (a) $\Wi=0.5$, (b) $\Wi=7$,
(c) $\Wi=20$.
Panel (d) shows the correlation times $T_{\hat{\bm R}}$ (black squares),
$T_{\hat{\bm z}_1}$ (blue circles),
$T_{\hat{\bm z}_1^\p}$ (red triangles) rescaled by $\tau_\eta$
as a function of the Weissenberg number.}
\label{fig:corr-orientation}
\end{figure}

In a coupled simulation of turbulent drag reduction, the feedback of polymers 
on the flow is of a tensorial nature; therefore, it depends not only on the extension
of polymers but also on their orientation.

For small values of $\Wi$, the separation vector $\bm R$ exhibits a weak alignment
with the eigendirections of the rate-of-strain tensor or with the vorticity,~\cite{WG10}
because thermal fluctuations (which are isotropic) strongly affect the dynamics.
In contrast,
for intermediate or large values of $\Wi$, thermal fluctuations have a negligible
effect on the dynamics of polymers. Consequently, the evolution of 
the extension $R$ decouples
from that of the orientation $\hat{\bm R}=\bm R/R$, which behaves like the orientation vector of
a rigid rod and is the solution of the Jeffery equation.~\cite{ORL82} For $Wi\gtrsim 5$, the statistics of
$\hat{\bm R}$ is therefore independent of $\Wi$ and coincides with that of the
orientation of a rigid rod, i.e. $\hat{\bm R}$ exhibits a moderate 
alignment with $\bm e_2$
and a strong alignment with $\hat{\bm \omega}$ (see
Refs.~\onlinecite{JC07,WG10,PW11} and Fig.~\ref{fig:pdf-orientation}).

In the FENE-P model, the first eigenvector of $\tC$, i.e. $\bm z_1^{\mathrm{P}}$, can be interpreted 
as the orientation of the polymer, provided $\Wi$ is sufficiently large.
The effect of the Peterlin approximation
on polymer orientation can then be studied by comparing the statistics of $\bm z_1^{\mathrm{P}}$
with that of the first eigenvector of $\C$.
Fig.~\ref{fig:pdf-orientation} shows that
the FENE-P model qualitatively reproduces the statistics of polymer orientation but underestimates the level
of alignment.

The autocorrelation function of $\hat{\bm R}$ decays exponentially 
(Figs.~\ref{fig:corr-orientation}(a) to~\ref{fig:corr-orientation}(c)),
which
for large $\Wi$ is in agreement with the results for the autocorrelation 
of the orientation of a rod.~\cite{PW11} The correlation time of
$\hat{\bm R}$ increases for small values of $\Wi$ and saturates at 
large $\Wi$ (Fig.~\ref{fig:corr-orientation}(d)), because
for large values of the Weissenberg number the dynamics of the orientation of $\bm R$ becomes
independent of $\Wi$. The FENE-P model agrees with the FENE model at small $\Wi$,
but quantitative discrepancies appear at large $\Wi$
(Figs.~\ref{fig:corr-orientation}(a) to~\ref{fig:corr-orientation}(c)).
In particular, the FENE-P model
underestimates the correlation time of polymer orientation
(Figs.~\ref{fig:corr-orientation}(d).

\section{Conclusions}
\label{sec:conclusions}

Numerical simulations of turbulent flows of polymer solutions use the FENE-P model,
which is based on the elastic dumbbell model but
requires a closure approximation for the elastic term. We have examined the
effect of the Peterlin approximation on the steady-state statistics of the extension and the orientation
of polymers. The FENE-P model captures the qualitative properties of the statistics,
but for large $\Wi$ it underestimates the steady-state probability of small extensions and
overestimates the probability of large extensions. 
As a consequence, the Peterlin approximation yields a greater 
average extension as well as a greater probability that
polymers break under the action of a turbulent flow.~\cite{T03}
To quantify this effect, one would need to couple the dynamics of polymers with a fragmentation 
model connected to the accumulated (or instantaneous) stress along each trajectory and to estimate 
the relative breaking rate.~\cite{BBL12}
Since large polymer extensions are more likely in the FENE-P model
than in the FENE model, we also expect that the Peterlin approximation yields
a stronger feedback of polymers on the flow in two-way coupling simulations of homogeneous
isotropic turbulence with polymer additives.
A similar argument, however, does not carry over to inhomogeneous flows like channel flows.
In this case, indeed,
drag reduction is caused by the strong stretching of polymers
in the near wall region rather than by the dynamics in the bulk of the channel, where the flow is homogeneous and 
isotropic and a lower degree of stretching is observed.\cite{BLP08,IDKCS02,ZA03,TDMS03,TDMSL04,BMPB12}

As regards the temporal statistics of the end-to-end separation vector,  
both the correlation times of the extension and the orientation of polymers are underestimated
by the FENE-P model.
The FENE-P model also underestimates the level of alignment of polymers
with the eigenvectors of the rate-of-strain tensor and with the direction of vorticity.

It would be interesting to check to what extent these properties of the FENE-P model
influence the dynamics of a polymer solution by comparing two-way coupling simulations 
of the FENE-P model in which the stress tensor is either calculated according to the 
Peterlin approximation or from molecular dynamics.

In this paper, we examined the Peterlin approximation, as this is the main assumption
in the construction of a continuum model of polymer solutions.
However, it is worth recalling that the FENE-P model is based on the dumbbell model and hence 
on a very simplified coarse-grained description of a polymer macromolecule.
Other approximations may thus impact the performance of the FENE-P model and its comparison
with experiments.
For instance, even for simple laminar flows, the dumbbell model reproduces the 
experimental observations only if the maximum extension and the effective bead radius
are used as free parameters to fit the experimental data.\cite{PSC97,LPSC97}

\acknowledgments This work was supported in part by the EU COST Action
MP1305 ``Flowing Matter'' and by a
research program of the Foundation for Fundamental Research on Matter
(FOM), which is part of the Netherlands Organisation for Scientific
Research (NWO).
Part of the computations were done at ``M\'esocentre
SIGAMM,'' Observatoire de la C\^ote d'Azur, Nice, France.
DV acknowledges the hospitality of the Department of Physics of the
Eindhoven University of Technology, where part of this work was done, and 
the Indo--French Centre for Applied Mathematics (IFCAM), Bangalore, for
financial support.

\end{document}